\documentclass[letterpaper,twocolumn,twocolappendix,appendixfloats]{emulateapj}
\usepackage{float}
\usepackage{epsfig}
\usepackage{graphicx}
\usepackage{graphics}
\usepackage[latin1]{inputenc}
\usepackage{latexsym}
\newcommand{\nd}{\multicolumn{1}{c}{$\dots$}}

\slugcomment{Accepted for publication in the Astronomical Journal}
\shorttitle{Photometry from Dome A}
\shortauthors{Wang et al.}

\begin{document}

\title{Photometry of Variable Stars from Dome A, Antarctica}
%\subtitle{Results of the 2008 Observing Season}

\author{Lingzhi Wang\altaffilmark{1,2}, Lucas M.~Macri,\altaffilmark{2}, Kevin
  Krisciunas\altaffilmark{2}, Lifan Wang\altaffilmark{2,3}, Michael
  C.~B.~Ashley\altaffilmark{4},\\Xiangqun Cui\altaffilmark{5}, Long-Long
  Feng\altaffilmark{3}, Xuefei Gong\altaffilmark{5}, Jon
  S.~Lawrence\altaffilmark{4,6}, Qiang Liu\altaffilmark{7}, Daniel
  Luong-Van\altaffilmark{4},\\Carl R.~Pennypacker\altaffilmark{8}, Zhaohui
  Shang\altaffilmark{9}, John W.~V.~Storey\altaffilmark{4}, Huigen
  Yang\altaffilmark{10}, Ji Yang\altaffilmark{3},\\Xiangyan
  Yuan\altaffilmark{5,11}, Donald G.~York\altaffilmark{12}, Xu
  Zhou\altaffilmark{7,11}, Zhenxi Zhu\altaffilmark{3} \& Zonghong
  Zhu\altaffilmark{1}}

\altaffiltext{1}{Department of Astronomy, Beijing Normal University, Beijing,
  100875, China}

\altaffiltext{2}{Mitchell Institute for Fundamental Physics \& Astronomy,
  Department of Physics \& Astronomy, Texas A\&M University, College Station,
  TX 77843, USA}

\altaffiltext{3}{Purple Mountain Observatory, Chinese Academy of
  Sciences, Nanjing 210008, China}

\altaffiltext{4}{School of Physics, University of New South Wales, NSW 2052,
  Australia}

\altaffiltext{5}{Nanjing Institute of Astronomical Optics and Technology,
  Nanjing 210042, China}

\altaffiltext{6}{Australian Astronomical Observatory, NSW 1710, Australia}

\altaffiltext{7}{National Astronomical Observatory of China, Chinese Academy of
  Sciences, Beijing 100012, China}

\altaffiltext{8}{Center for Astrophysics, Lawrence Berkeley National
  Laboratory, Berkeley, CA, USA}

\altaffiltext{9}{Tianjin Normal University, Tianjin 300074, China}

\altaffiltext{10}{Polar Research Institute of China, Pudong, Shanghai 200136,
  China}

\altaffiltext{11}{Chinese Center for Antarctic Astronomy, Nanjing 210008,
  China}

\altaffiltext{12}{Department of Astronomy and Astrophysics and Enrico Fermi
  Institute, University of Chicago, Chicago, IL 60637, USA}

\begin{abstract}

  Dome A on the Antarctic plateau is likely one of the best observing sites on
  Earth thanks to the excellent atmospheric conditions present at the site
  during the long polar winter night. We present high-cadence time-series
  aperture photometry of 10,000 stars with $i<14.5$~mag located in a 23
  square-degree region centered on the south celestial pole. The photometry was
  obtained with one of the CSTAR telescopes during 128 days of the 2008
  Antarctic winter.

  We used this photometric data set to derive site statistics for Dome A and to
  search for variable stars. Thanks to the nearly-uninterrupted synoptic
  coverage, we find $6\times$ as many variables as previous surveys with
  similar magnitude limits. We detected 157 variable stars, of which 55\% are
  unclassified, 27\% are likely binaries and 17\% are likely pulsating
  stars. The latter category includes $\delta$~Scuti, $\gamma$ Doradus and RR
  Lyrae variables. One variable may be a transiting exoplanet.

\end{abstract}


\section{Introduction}

Long, continuous, unbroken time-series photometry is highly advantageous for a
range of astrophysical problems, such as exoplanet searches, studies of periodic
variable stars and discoveries of previously-unknown stellar behavior. While
these data can be obtained via coordinated observations by a world-wide
telescope network, such an approach is fraught with calibration issues, variable
weather across the sites, and is highly labor intensive. Ideally, the data would
be acquired with a single, fully autonomous robotic telescope.  Space is
obviously an ideal place to acquire such data due to the high photometric
quality and long observing sequences that can be achieved, as demonstrated by
the CoRoT \citep{BaglinA2006cosp-CoRoT,BoisnardL2006ESASP-CoRoTInBrief} and
Kepler missions \citep{Benko2010,Borucki2010}. Antarctica offers the only
location on the surface of the earth where similar observations can be obtained.

\ \par
 
The Antarctic plateau offers an unparalleled opportunity to make these types
of observations with a single telescope. The combination of high altitude, low
temperature, low absolute humidity, low wind and extremely stable atmosphere
opens new windows in the infrared and terahertz regions and offers improved
conditions at optical and other wavelengths. These advantages offer astronomers
gains in sensitivity and measurement precision that can exceed two orders of
magnitude over even the best temperate sites
\citep{Storey2005AS-AstronomyFromAntarctica,
  Storey2007CAA-Antarctic,Storey2009AAPPS-AAA,Burton2010AAR-Antarctic}. Additionally,
the high duty cycle and long duration of observations from Antarctic sites such
as Dome A and Dome C allow for efficient asteroseismic
observations. Observations with one site on the Antarctic plateau offers a
performance similar to or better than a six-site network at other latitudes
\citep{Mosser2007PASP-DutyCycle}.

\ \par

Furthermore, sites on the Antarctic plateau suffer from less high-altitude 
turbulence than temperate sites and thus enjoy lower scintillation noise 
leading to superior photometric precision \citep{KenyonSL2006PASP-ScintillationDomeC}. 
The elevation of a source observed from Antarctica changes little during the
the course of 24 hours, again improving photometric precision and providing the 
potential for long, continuous time-series observations. 

\ \par

Other advantages include dark skies and lower precipitable water vapor
(PWV) which leads to better atmospheric transmission. Some
disadvantages that must be considered include aurorae, a reduced
amount of the celestial sphere that is available for observations, and
prolonged twilight. One site on Antarctica, Dome C, has a similar
number of cloud-free astronomically dark time and has lower
atmospheric scattering than low-latitude sites, reducing the sky
brightness and extinction \citep{KenyonSL2006PASP-OpticalSkyBrightnessExtinction}.

\ \par

A very promising site on the Antarctic plateau is Dome A, at an elevation of
4,093m (840m higher than Dome C), where the Chinese Kunlun station is under
construction. Dome A is possibly the best astronomical site on Earth based on
data, models, and meteorological parameters such as cloud cover, boundary layer
characteristics, aurorae, airglow, free atmosphere, and precipitable water vapor
 \citep{Saunders2010EAS-BestSite,Saunders2009PASP-BestSite}.
\citet{SwainMR2006PASP-AntarcticBoundaryLayerSeeing} have modeled the turbulent
surface layer across the entire Antarctic ice sheet and found thickness
differences from site to site, with one of the thinnest regions being at Dome
A. This was confirmed by Snodar measurements in 2009. The median thickness of
the boundary layer was 14m \citep{Bonner2010PASP-TBL}.

\vspace{6pt}

Long-term time-series photometric data has been acquired at the South Pole and
at Dome C. The 5-cm South Pole Optical Telescope was used to obtain continuous
observations over 78 hours of the Wolf-Rayet star $\gamma^{2}$
Velorum \citep{TaylorM1990AJ-PhotometryWolfRayetStar}.
\citet{StrassmeierKG2008AA-FirstOpticalPhotometryDomeC} used the 25-cm sIRAIT
telescope to obtain continuous (98\% duty cycle over 10 days) photometry of two
bright variable stars. The resulting light curves had {\it rms} scatter of 3
mmag in the V band and 4 mmag in the R band over a period of 2.4hrs, values that
are $3-4$ times better than previously obtained with an equivalent telescope in
southern Arizona. They attributed the improved photometric precision to the
exceptionally low scintillation noise in the Antarctic plateau.  The ASTEP
project \citep{CrouzetN2010AA-ASTEPSouth} used a fixed 10 cm refractor to
monitor a 15-sq degree field centered on the South Celestial Pole during 4
months of operation (2008 June-September). They found sky conditions suitable
for observations were present for $\sim 85\%$ of the time and calculated
an observing efficiency for the detection of short-period exoplanets around
bright stars that is a factor of $1.5\times$ higher than that of temperate sites. 

\vspace{6pt}

An observatory that can operate year round without interruptions is required to
best capitalize upon the advantages provided by the Antarctic plateau. We have
built such an observatory at Dome A, called PLATO \citep{Ashley2010HiA-PLATO,
  Luongvan2010SPIE-PLATO, Yang2009PASP-PLATO,Lawrence2009RSI-PLATO,
  Lawrence2008SPIE-PLATO, Hengst2008SPIE-PLATO,Lawrence2006SPIE-SiteTestA}, and
a quad-telescope called CSTAR \citep[the Chinese Small Telescope
ARray,][]{Yuan2008SPIE-CSTAR,Zhou2010RAA-Testing}. Based on a large amount of
high-quality photometric data obtained during the 2008 Antarctic winter,
\citet{Zou2010AJ-Sky} undertook a variety of sky brightness, transparency and
photometric monitoring observations, while \citet{Zhou2010PASP-Catalog}
published a catalog of $\sim10,000$~stars in a field centered on the South
Celestial Pole.

\vspace{6pt}

This paper presents an independent analysis of the data acquired by CSTAR during
the 2008 Antarctic winter season. \S2 briefly describes the instrument, observations and
data reduction; \S3 presents details of the photometric procedure and the
astrometric and photometric calibrations; \S4 describes the steps followed to
obtain high-precision time-series photometry of the brightest 10,000 stars; \S5
presents a catalog of variable stars and rough statistics of variable star
types; \S6 contains our conclusions.

\vfill
 
\section{Observations and data reduction}

\subsection{Observations} 

CSTAR \citep{Yuan2008SPIE-CSTAR,Zhou2010RAA-Testing}, a part of the PLATO
observatory\footnote{http://mcba11.phys.unsw.edu.au/$\sim$plato/cstar.html}, is
the first photometric instrument deployed at Dome A. It was built by astronomers
Xu Zhou and Zhenxi Zhu and is composed of four Schmidt-Cassegrain wide field
telescopes. Each CSTAR telescope has a field of view $4.5^{\circ}$ in diameter
with a pupil entrance aperture of 145 mm. Each focal plane contains an ANDOR
DV435 $1K\times1K$ frame-transfer CCD with a pixel size of 13 $\mu$m, which
translates to a plate scale of $15\arcsec/$~pix. Three of the telescopes have
$g$, $r$, and $i$ filters similar to those used by Sloan Digital Sky Survey
\citep{Fukugita1996}. Table 1 of \citet{Zhou2010RAA-Testing} lists the effective
wavelength (470nm, 630nm, 780nm) and full width half maximum (140nm, 140nm,
160nm) of the three filters ($g$, $r$, $i$).  No filter is used in the fourth
telescope.  The field of view of the telescopes is very nearly centered on the
South Celestial Pole. Since the telescopes are fixed, stars move in circles
about the center of the CCD but only traverse a tiny fraction of a pixel during
a single exposure. More details on CSTAR can be found in
\citet{Zhou2010PASP-Catalog,Zhou2010RAA-Testing}.

The CSTAR telescopes were installed at Dome A in January 2008. During the
following Antarctic winter season, there were technical problems with 3 of the
4 telescopes -- those that use the Sloan $g$, Sloan $r$, and ``open'' filters
-- which prevented them from obtaining any useful data. Fortunately, the fourth
telescope (equipped with the Sloan $i$ filter) performed without any
significant issues. Observations were conducted from 2008 March 20 through 2008
July 27; during this interval, more than 287,800 frames were acquired with a
total integration time of 1,615 hours. The total amount of raw data collected
during the observing season was about 350 GB. Table~\ref{tb:log} lists the
number of images acquired and total exposure time per month, while
Table~\ref{tb:exp} details the different exposure times used during the
observing season.

Two groups have carried out independent analyses of the data; one at the
National Astronomical Observatories of the Chinese Academy of Sciences 
\citet{Zhou2010PASP-Catalog} and another at Texas A\&M University and Beijing
Normal University (present work).  A comparison of the photometric precision of
the two reductions is given in \S\ref{sec:prec}.

\subsection{Data Reduction}

The preliminary reduction for the raw science images involved bias subtraction,
flat fielding, and fringe correction. We used a bias frame obtained during
instrument testing in China. We created a sky flat by median combining 160
images with high sky levels ($\geq$ 15,000 ADU) obtained on 2008 March 17.

The bias-subtracted and flat-fielded $i$-band images contained a residual
fringe pattern with peak-to-peak variations of $\sim1.4$\% of the sky value. The
fringes are due to variations in the thickness of the CCD and strong emission
lines in this region of the visible spectrum. An important aspect of these
features is that they introduce an additive contribution to the signal and they
should be subtracted to perform photometrically consistent measurements.

\begin{deluxetable}{lrr}
\tablewidth{0pt}
\tablenum{1}
\tablecaption{Log of observations\label{tb:log}}
\tablehead{\colhead{Month} & \colhead{\# images} &
\colhead{Total exp.}\\ \colhead{2008} & & \colhead{time (hr)}}
\startdata
March       &  14345 &   64.4 \\
April       &  51390 &  316.4 \\
May         &  71665 &  398.1 \\
June        & 110358 &  613.1 \\
July        &  40082 &  222.7 \\
{\bf Total} & 287840 & 1614.7
\enddata
\end{deluxetable}

\addtocounter{table}{1}
\begin{deluxetable}{llll}
\tablewidth{0pt}
\tablenum{2}
\tablecaption{Exposure times\label{tb:exp}}
\tablehead{\colhead{Start} & \colhead{End} & \colhead{Exp.} & \colhead{\#}\\
\colhead{date} & \colhead{date} & \colhead{(s)} & \colhead{images}}
\startdata
2008-03-20 & 2008-03-23 &  5 &   7944 \\
2008-03-23 & 2008-04-11 & 30 &  17514 \\
2008-04-11 & 2008-07-27 & 20 & 262382
\enddata
\end{deluxetable}

We created a fringe correction image by carrying out the following two-step process.
We selected 3,450 bias-subtracted and flat-fielded images obtained during 2008
March which exhibited a strong fringe pattern. We first masked all saturated
pixels (values above 25,000 ADU) and their nearest neighbors. We then carried
out PSF photometry using an automated pipeline based on DAOPHOT and ALLSTAR 
\citep{Stetson1987}. We set the detection threshold to $5\sigma$ above sky. We
masked the pixels associated with every star detected in each frame, using
different radii for different ranges of instrumental magnitude (15 pix for
$m<15$~mag, 10 pix for $15<m<17$~mag and 7 pix for $m>17$~mag). All remaining
pixels on the masked images represented contributions from the sky background
or stars below our detection threshold (see below). We grouped the images into
23 sets of 150 frames each, placing consecutive images in different sets to
minimize the overlap of masked regions. Since the CSTAR telescopes do not track
the rotation of the sky, interleaving consecutive images into different sets
ensures that the pixels masked due to the presence of stars will fall at
different positions on the detector. We normalized every image by its median
sky value and median combined each set. We then median combined the 23
intermediate images into an initial fringe correction image.

The initial fringe correction image could be affected by faint stars below the
$5\sigma$ detection threshold used in the above procedure. In order to remove
such objects, we applied a preliminary fringe correction to the input frames
and lowered the detection threshold to $2.5\sigma$ above sky. We merged the
masks obtained in each of the two steps and repeated the combination process
described above to obtain a final fringe correction image.

The fringe correction image was subtracted from each science-quality frame in
an iterative manner, scaling its amplitude until the residuals in two corner
areas of the frame were statistically indistinguishable.

%scaling it by the estimated sky value until the residuals in two corner areas
%of the frame ([100:300,700:900] and [700:900,100:300]) were statistically
%indistinguishable.

\section{Photometry}\label{sec:zpt}

\subsection{Frame selection and photometry}

\begin{figure}[t]%sun elevation angle
\begin{center}
\includegraphics[width=0.45\textwidth]{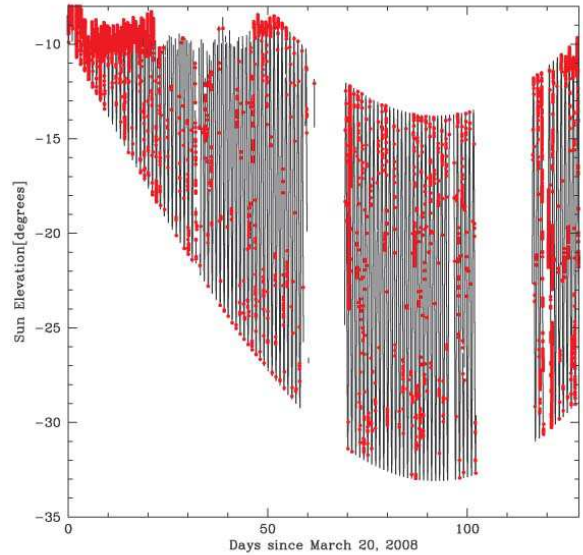}
\caption{Sun elevation angle for each science-quality frame obtained with CSTAR
  during the Antarctic winter of 2008. Black points indicate frames suitable for
  analysis, while red points denote frames that failed to meet our selection
  criteria.\label{fig:sun}}
\end{center}
\end{figure}

We performed photometry on the debiased, flattened, fringe-corrected images
using a pipeline based on DAOPHOT and ALLSTAR. We used the FIND and SKY routines
in the IDL version of DAOPHOT to reject images taken under very high sky
background or cloudy conditions. We selected frames with a sky level below 6,000
ADU and more than 1,500 stars in the frame; these criteria were met by 93\% of
the images (about 270,000 frames). Figure \ref{fig:sun} shows the distribution
of the selected and rejected frames as a function of date and sun elevation
angle. 49\% of the data obtained in March was rejected due to the high sky
background caused by the relatively large sun elevation angle. Only 4\%, 5\%,
1\% and 11\% of the data acquired in April, May, June and July, respectively,
were rejected. Based on Fig.~\ref{fig:sun}, a suitable initial cut for useful
$i$-band data at Dome A would be a solar elevation angle below
$-10^{\circ}$. 96\% of all frames obtained below that value passed our
selection criteria.

\begin{figure}[t]%master image
\begin{center}
\includegraphics[width=0.45\textwidth]{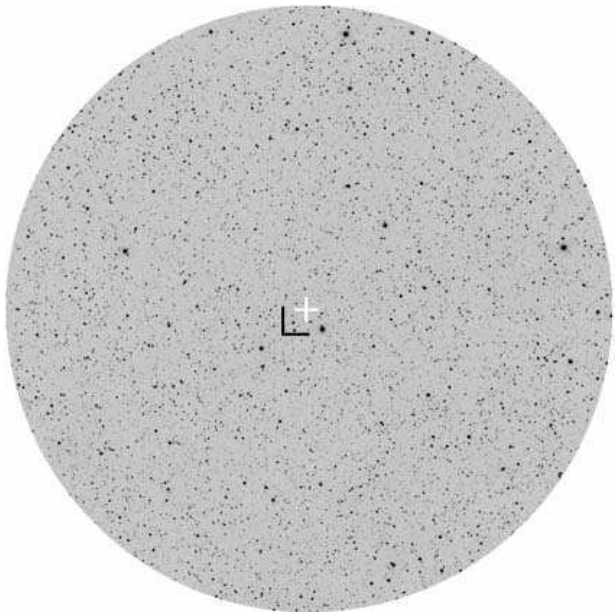}
\caption{Master frame of the CSTAR field of view. The white cross
  denotes the position of the South Celestial Pole. The edge of the
  field corresponds to $\delta=-87^{\circ}17\arcmin$. The black lines
  indicate the bottom-left corner of the inset image shown in
  Fig.~\ref{fig:radprof}.\label{fig:master}}
\end{center}
\end{figure}

Frame-by-frame registration was used to combine 3000 of the best images obtained
during a 24-hour period of exceptionally good conditions (2008 June 29) to form
a master reference image. We selected images with sky values below 500 ADU and
with more than 10,000 stars above the detection threshold. We combined the
images using MONTAGE (developed by P.~Stetson) into a master image that was
resampled (via bicubic interpolation) at $4\times$ the initial pixel scale. We
restricted the master frame to a circular field of view that was observed
continuously, spanning $-90^\circ < \delta < -87^\circ17\arcmin$.  We masked
saturated stars and their surroundings using a circular mask of varying
radius. We identified all stars in the masked master image using DAOFIND and a
detection threshold of $5\sigma$. The master image is shown in
Fig.~\ref{fig:master}.

\begin{figure}[t]%master image zoom
\begin{center}
\includegraphics[width=0.45\textwidth]{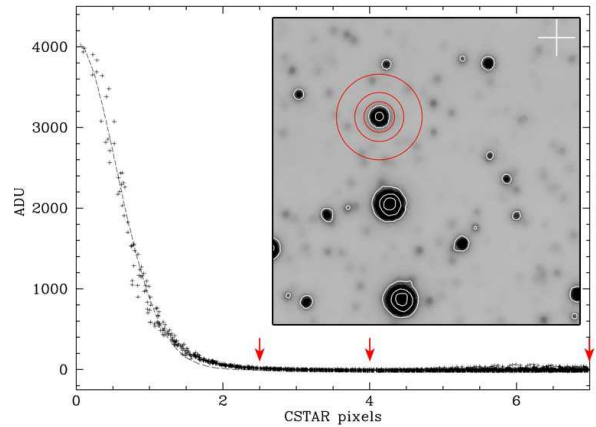}
\caption{Inset: Zoom into a small region of the master image of the
  CSTAR field ($12\arcmin$ on a side) near the South Celestial Pole
  (indicated by the white cross as in Fig.~\ref{fig:master}). The
  white contours indicate intensity levels logarithmically spaced
  between 20 and $10^4$ ADU. The red circles indicate the extent of
  the aperture radius and the sky annulus, centered on a random
  star. Outer plot: radial profile for the same star, indicating the
  location of the aperture radius and sky annulus.\label{fig:radprof}}
\end{center}
\end{figure}

Next, we carried out aperture and point-spread function (PSF) photometry using
the stand-alone versions of DAOPHOT and ALLSTAR. The aperture photometry radius
was set to 2.5 pixels, with the sky annulus extending from 4 to 7
pixels. Fig.~\ref{fig:radprof} shows the radial profile and the
location of these radii for a random star.

\begin{figure}[t]%comparison ap and psf photometry
\begin{center}
\includegraphics[width=0.45\textwidth]{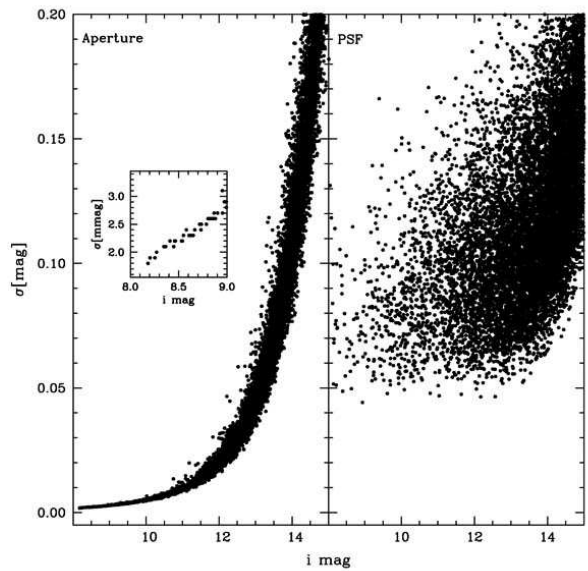}
\caption{Comparison of DAOPHOT/ALLSTAR uncertainties for aperture (left) and
  PSF (right) magnitudes for a typical frame. The inset zooms in the aperture
  photometry for $8 < m < 9$. \label{fig:appsf}}
\end{center}
\end{figure}

The smallest uncertainties reported by DAOPHOT for aperture
photometry were of the order of 2~mmag. We modeled the PSF using a Moffat
function with $\beta=1.5$, which gave the smallest residuals. Due to the severe
undersampling of the camera ($15\arcsec/$pix), the smallest uncertainties
reported by ALLSTAR for PSF photometry were significantly larger ($\sim
0.05$~mag) than those reported for aperture photometry. Fig.~\ref{fig:appsf}
shows a comparison between the reported DAOPHOT aperture and PSF photometric
uncertainties. We conclude that PSF photometry is not suitable for the analysis
of CSTAR data, and we only use aperture photometry for the
remainder of this paper.

\begin{figure}[t]%info plot:extinction, sky level ,N stars statistics
\begin{center}
\includegraphics[width=0.45\textwidth]{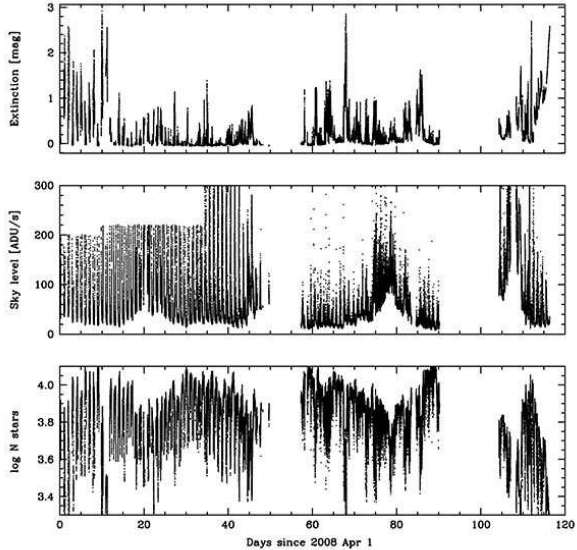}
\caption{Time series plots of differential extinction (top panel), sky brightness 
level (middle panel), and number of stars on a single image (bottom panel).
\label{fig:info}}
\end{center}
\end{figure}

\begin{figure}[t]
\begin{center}
\includegraphics[width=0.45\textwidth]{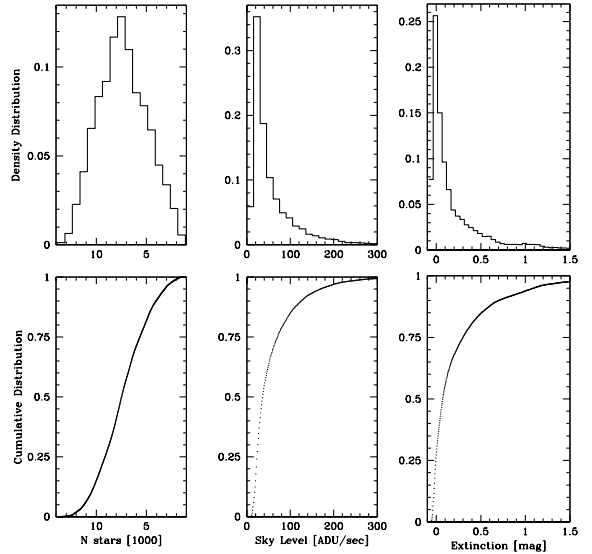}
\caption{Frequency of occurrence and cumulative fraction of the number of stars
  per image (left column), sky brightness level (middle column), and
  differential extinction (right column). \label{fig:chist}}
\end{center}
\end{figure}

We carried out aperture photometry on selected scientific images to determine
for each one the number of stars, the sky level, and the photometric zeropoint
(proxy for extinction due to clouds) relative to the master frame. The latter
was calculated from the magnitude offset of the brightest 1,000 stars in each
image across the whole field of view with respect to the corresponding
magnitudes in an individual image chosen as our internal photometric
reference. Outlier rejections of 3$\sigma$ were applied iteratively to guarantee
the zeropoints were based on non-variable stars. Fig.~\ref{fig:info} shows the
time series of number of stars, sky level and photometric zeropoint (proxy for
differential extinction due to clouds), while Fig.~\ref{fig:chist} shows the
frequencies of occurrence and cumulative density distributions of the same
quantities.

There is a small zeropoint offset of $\sim0.03$~mag between our
reference image and the one used by \citet{Zou2010AJ-Sky}. Hence, the
peak value of the distribution in the top right panel in
Fig.~\ref{fig:chist} is shifted to the left by that amount relative to
Fig.~5 of \citet{Zou2010AJ-Sky}. Based on these values, we can
conclude that extinction due to clouds in the $i$ band at Dome A is
less than 0.4~mag during 80\% of the time and less than 0.1~mag for
50\% of the time. Our results are consistent with those presented in
Table~1 of \citet{Zou2010AJ-Sky}.

50\% of the science-quality images have at least 7,500 stars above the
detection threshold and a sky level below 36 ADU/s. The latter is equivalent to
a median sky background of 19.6 mag$/\sq \arcsec$, consistent with the
value of 19.8 mag$/\sq \arcsec$ derived by \citet{Zou2010AJ-Sky}. That paper also
compared the median sky background from La Palma, Paranal, Cerro Tololo, and
Calar Alto and concluded that under moonless clear conditions Dome A has darker
sky backgrounds than the above astronomical sites. 

\vfill

\subsection{Astrometric and Photometric calibration}\label{sec:calib}

We determined the astrometric solution of our master reference image based on
an independent astrometric calibration of a single frame (15CC0007) obtained
during the same observing season by M.~Ashley (priv. comm.). We found 5,300
stars in common between this frame and our master reference image using
DAOMATCH and DAOMASTER (developed by P.~Stetson). We extracted the celestial
coordinates of these stars from the calibrated frame using the ``xy2sky''
routine in WCSTools\footnote{http://tdc-www.harvard.edu/wcstools} and
determined the astrometric solution of our master reference image using the
routines ``ccmap'' and ``ccsetwcs'' in IRAF\footnote{IRAF is distributed by the
  National Optical Astronomy Observatory, which is operated by the Association
  of Universities for Research in Astronomy, Inc., under cooperative agreement
  with the National Science Foundation (NSF).}. ccmap was used to compute plate
solutions using matched pixel and celestial coordinate lists and ccsetwcs was
used to create an image world coordinate system from a plate solution. We used
the TNX projection for the WCS of our master reference image.

\begin{figure}[t]%astrometry calibration plot
\begin{center}
\includegraphics[width=0.45\textwidth]{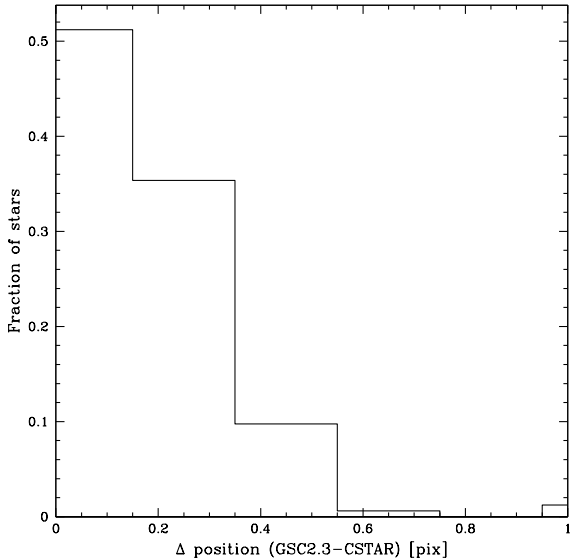}
\caption{Distribution of astrometric differences between CSTAR and the
 Guide Star Catalog v2.3. The plate scale of CSTAR is
 $15\arcsec/pix$.\label{fig:AstrometryErr}}
\end{center}
\end{figure}

We determined a mean astrometry uncertainty of $\sim5\arcsec$ ($\sim 1/3$~pix)
by comparing the coordinates we derived and those tabulated in the Guide Star
Catalog \citep[version 2.3,][]{Lasker2008} for selected stars across the entire
field of view of our master frame.  Fig.~\ref{fig:AstrometryErr} shows the
distribution of positional differences for these stars. 

\begin{figure}[t]%zpt plott
\begin{center}
\includegraphics[width=0.45\textwidth]{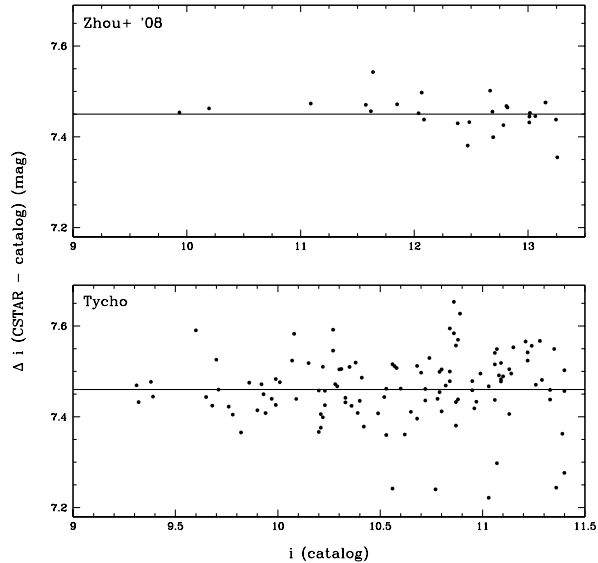}
\caption{Photometric calibration of the CSTAR observations.\label{fig:zpt}}
\end{center}
\end{figure}

We determined the photometric calibration of our master reference image using
the catalog of calibrated $griz$ magnitudes of Tycho stars \citep{Ofek2008}.
We used Vizier\footnote{http://vizier.u-strasbg.fr} to select stars with
$9<i<11.5$~mag and $\delta < -87^\circ$ and matched them to objects in our
master image using a positional tolerance of 2.5 pixels ($=37.5\arcsec$). We
applied an iterative outlier rejection and determined a photometric zeropoint
of $7.46\pm0.08$~mag based on $N=119$~stars, shown in the bottom panel of
Fig.~\ref{fig:zpt}.

We checked the above calibration by matching stars in our master frame with
bright unblended stars in the catalog of \citet{Zhou2010PASP-Catalog}, who
computed aperture photometry using a 3-pixel radius. Those authors adopted a
photometric calibration based on 48 stars in common with the USNO-B catalog,
whose $i$-band magnitudes were determined by \citet{Monet2003AJ-USNOB}. We
determined a photometric zeropoint of $7.45\pm0.04$~mag (see top panel of
Fig.~\ref{fig:zpt}), which is statistically identical to the value determined
above.We adopt the calibration based on the Tycho Catalog, as it is based on a
larger number of stars.

\section{Time-series photometry}

There are approximately 100,000 stars detected in our master reference image,
reaching a depth of $i=20.4$~mag. The brightest 10,000 of these are detected in
most individual frames and correspond to the depth of a previous study of this
area of the sky by the ASAS project \citep[$V\sim14.5$~mag][]{Pojmanski2005}.
Hence, we selected these objects for time-series photometry with the aim of
detecting variable stars and determining the most significant frequencies in
their power spectra. Hereafter, we will refer to this subset as the
``bright-star sample''. We restricted our analysis to the images obtained under
the best conditions, defined as a sky background below 50 ADU/s and extinction
$\leq0.5$~mag. These conditions were met by 141,700 frames or about 53\%
of all the science-quality data.

\subsection{Photometric corrections\label{sub:photcor}}

We performed a series of corrections to the aperture photometry reported by
DAOPHOT, which included: exposure time normalization, zeropoint correction,
sigma rescaling, residual flat fielding, time calibration, masking of satellite
trails and saturated regions, spike filtering, and magnitude calibration. In
greater detail, these corrections entailed the following steps:

\begin{itemize} 

\item Exposure time normalization: The aperture photometry reported by 
    DAOPHOT was corrected by the usual factor of $2.5 \log(t)$.

\item Zeropoint correction: The zeropoint difference between each image and the
  photometric reference image was determined following the procedure previously
  described in \S\ref{sec:zpt}

\item Sigma rescaling: We rescaled the magnitude uncertainties reported by 
    DAOPHOT so that $\chi_{\nu}^2=1$ for non-variable stars, following the
  procedure developed by \citet{Kaluzny1998AJ-RescaleDAOphotErr}.

\item Residual flat-fielding: We selected 200 bright non-variable stars and
  searched for correlations between their $(x,y)$ position and deviations from
  their mean magnitude on every science-quality frame. We averaged the results
  from neighboring pixels to increase $S/N$, at the cost of decreasing the
  spatial resolution by $4\times$. The result is equivalent to a residual
  flat-field frame, which exhibited peak-to-peak variations of 1\%.

\item Time calibration: The clock in the data acquisition computer exhibited a
  small drift during the observing season. We corrected the Julian Dates of all
  science-quality frames following \S4.3 of \citet{Zhou2010PASP-Catalog}.

\item Masking of satellite trails: Numerous satellite trails are present in the
  CSTAR images, easily identifiable as lines with fluxes $>10\sigma$ above
  background. This polluted stars whose positions on the CCD overlapped the
  trails. We used a parallelogram mask with the trail line as the bisector and
  a width that encompassed all pixels with fluxes $>3\sigma$ above
  background. This was equivalent to a width of 20 pixels or less, depending on
  the brightness of each trial. We rejected the individual photometric
  measurements of stars lying inside the masked regions of a given frame.
\end{itemize}

\vspace{-12pt}
\begin{itemize}
\item Masking of saturated regions: We masked pixels with values above 25,000
  ADU and their associated bleed trails.

\item Spike filtering: We detected quasi-periodic, short-duration (typically 30
  min) increases in the brightness of some stars (typically 0.4 mag) which were
  otherwise not variable. Some of these ``spikes'' were due to ghosts (multiple
  reflections of incoming photons) of bright stars which we identified by
  subtracting neighboring images. We initially identified spikes by searching
  for $3\sigma$ deviations from the mean magnitude in 6-hour blocks of data. In
  the case of periodic variable stars, the mean was computed over $\frac{1}{8}$
  of the period or 0.1 day, whichever was smaller. We carried out the procedure
  twice, shifting the starting time of each bin by half of its width on the
  second pass. This algorithm might have eliminated some {\em bona fide}
  aperiodic flux variations (such as stellar flares), but we defer a thorough
  analysis of such events and other causes of these spikes to a future
  publication.

\item Magnitude calibration: We placed our corrected instrumental magnitudes on
  the standard system as described previously in \S\ref{sec:calib}.
\end{itemize}

\subsection{Photometric Precision}\label{sec:prec}

\begin{figure}[t]
\begin{center}
\includegraphics[width=0.45 \textwidth]{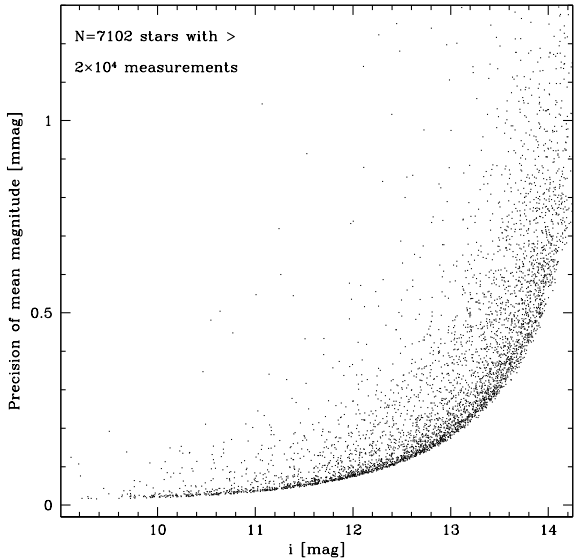}
\caption{Photometric precision of mean $i$ magnitudes for more
  than 7000 stars that have at least 20,000 observations.\label{fig:magerr}}
\end{center}
\end{figure}

Over 70\% of the bright-star sample considered in this analysis have more than
20,000 photometric measurements. Thanks to this, the internal statistical
uncertainty in the mean magnitudes of stars with $i\leq 13.5$~mag is less than
$10^{-3}$ mag, as seen in Fig.~\ref{fig:magerr}. As discussed previously in
\S\ref{sec:calib}, the {\it absolute} uncertainty of the photometry is limited
by the synthetic magnitude calibration using the Tycho catalog.

\begin{figure}[t]%yerror comparison with Zhou
\begin{center}
\includegraphics[width=0.45 \textwidth]{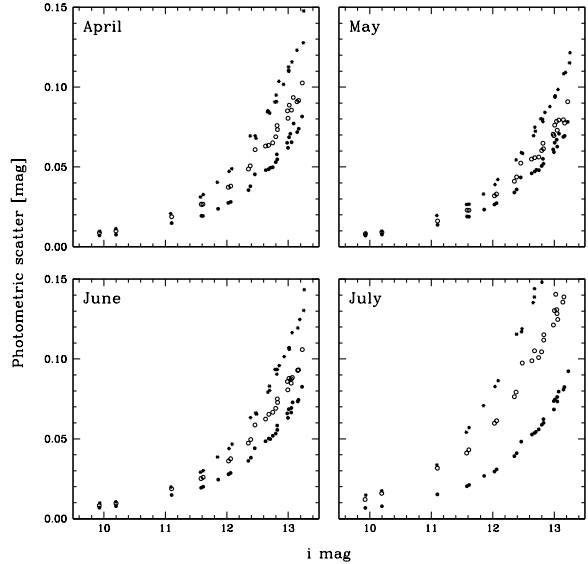}
\caption{Comparison of photometric scatter for stars in common with 
  \citet{Zhou2010PASP-Catalog} that were used to determine the zeropoint offset
  in the top panel of Fig.~\ref{fig:zpt}. Separate comparisons are show for
  April, May, June and July. Starred symbols: photometry from 
  \citet{Zhou2010PASP-Catalog}; open symbols: our aperture photometry before
  applying the photometric corrections described in \S\ref{sub:photcor}; filled
  symbols: our aperture photometry after applying the corrections.\label{fig:cmpzhou}}
\end{center}
\end{figure}

Fig.~\ref{fig:cmpzhou} compares the scatter in our mean magnitudes
with the results of \citet{Zhou2010PASP-Catalog}. We compared 27 stars
in common between our catalogs which were used to obtain the
alternative zeropoint discussed in \S\ref{sec:calib}.  We calculated
the scatter in the mean magnitude of each star in both data sets using
a $3\sigma$ iterative outlier rejection, performing separate
comparisons for each month of the observing season. Our mean
magnitudes exhibit a slightly smaller scatter than
\citet{Zhou2010PASP-Catalog}. Some possible reasons for the reduced
scatter include the use of a smaller sky annulus located closer to
each star and the rejection of measurements affected by satellite
trails.

\section{Variable Star Catalog and Statistics} 

\subsection{Search for Variability}

\begin{figure}[t]%L-statistics
\begin{center}
\includegraphics[width=0.45\textwidth]{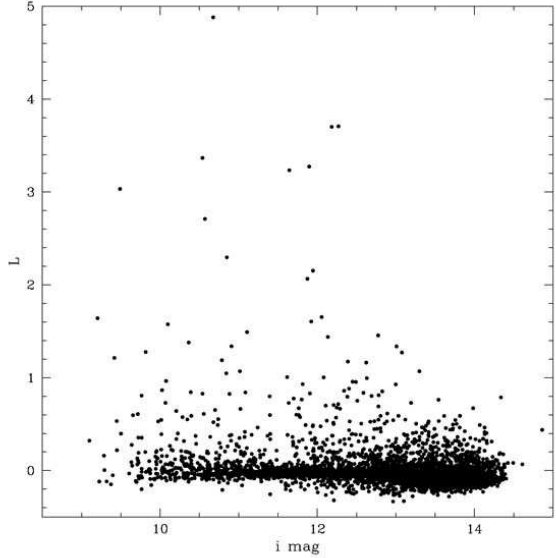}
\caption{Variability statistic $L$\ \citep{Stetson1996} for the 10,000
  brightest stars in the CSTAR sample.\label{fig:lste}}
\end{center}
\end{figure}

We applied several techniques to quantify the variability of the bright-star
sample. First, we computed the Welch-Stetson variability index $L$ \citep[\S2
  of][]{Stetson1996}, which measures the time-dependent correlation of
magnitude residuals of pairs of close observations, rescaling them according to
the estimated standard error of each magnitude measurement. Fig.~\ref{fig:lste}
shows the distribution of $L$ values as a function of magnitude for this
subsample.

We searched for periodic variability among the bright-star sample using three
techniques. The first technique is the Lomb-Scargle method \citep[][hereafter
  LS]{Lomb1976, Scargle1982}, which applies the statistical properties of
least-squares frequency analysis of unequally spaced data on a series of test
periods. We searched for periods between 0.1 and 50~d and used a bin size of
0.01~d. Periods with $S/N\geq 10$ in the periodogram were considered to be
significant. The second technique involves Fourier decomposition as implemented
by the Period04 program \citep{Lenz2005}, which is described in detail in
\S~\ref{sec:multper}. We searched for frequencies between 0 and 50 cycles/day
with a bin size set by the Nyquist sampling. Frequencies with $S/N\geq 4$ were
considered significant.

Lastly, we searched for periodic transit events on prewhitened light curves
using the box fitting algorithm \citep[][hereafter BLS]{Kovacs2002}. This
algorithm looks for signals characterized by a periodic alternation between two
discrete levels with much less time spent at the low-level (occultation) phase.
The duration of a transit was allowed to range between 0.01 and 0.1 of the
primary period.  The BLS transit period was sought over the same time span as
the LS period using 10,000 trial periods and 200 phase bins. We prewhitened the
selected light curves based on the most significant period ($1/f$) and its 10
higher-harmonics (frequencies of $2f$, $3f, \dots, 11f$) and 9 sub-harmonics
(frequencies of $f/2$, $f/3, \dots, f/10$)

We found 115 periodic variables by the LS method, 29 additional periodic
variables via the Fourier decomposition technique, and 10 transit events
through the BLS method, for a total of 154 periodic variables.

\begin{figure}[b]
\begin{center}
\includegraphics[width=0.45\textwidth]{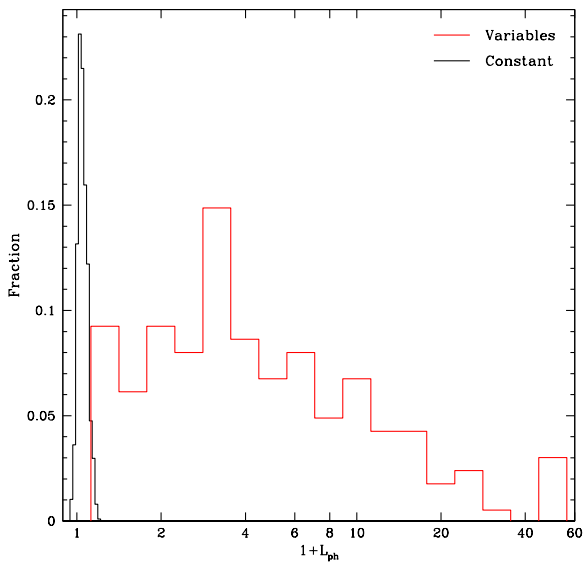}
\caption{Histogram of variability statistic $L$ for objects that exhibited a
  significant periodicity using the LS technique (red), compared to the
  equivalent distribution for objects without a significant periodicity
  (black). The variability statistic was computed using phased light curves
  with $10^3$ points per cycle.\label{fig:lstehist}}
\end{center}
\end{figure}

We phased the light curves of these stars using the most significant period and
calculated the median value for every $10^{-3}$ in phase. We calculated the $L$
values for the binned phased light curves of these stars and compared the
distribution to the one for 100 stars which had no significant periodicity.
Each star in the latter sample was phased using all periods found in the former
set, yielding $\sim15,000$ test light curves. Fig.~\ref{fig:lstehist} shows a
comparison of both distributions. We selected 149 out of 154 objects with
$L_{ph}> 0.25$ as our final sample of periodic variables. We further included 8
variables with significant changes in magnitudes but no periodicity within our
observing window (128 days). In these cases, we binned the data as a function
of time every 3000s and found values of $L_{ph}>2$ in all cases.

\begin{figure}[t]%unphased light curves
\begin{center}
\includegraphics[width=0.45\textwidth]{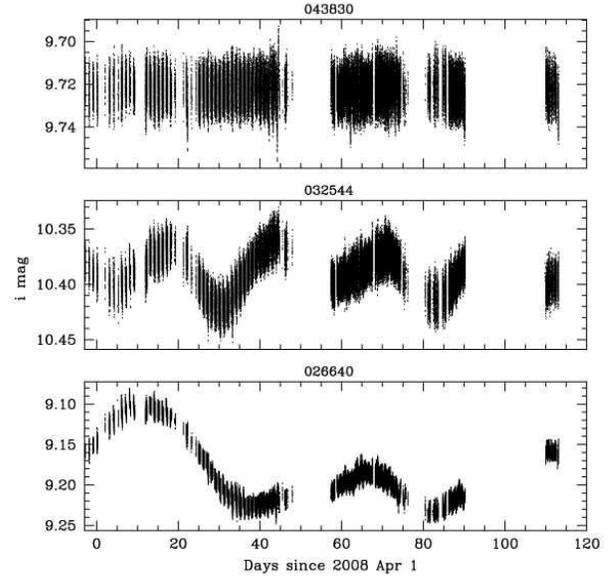}
\caption{Light curves of a bright constant star (top panel) and two variable
  stars (bottom two panels).\ \\ \label{fig:lcunph}}
\end{center}
\end{figure}

\begin{figure}[b]%lc phased lc plots
\begin{center}
\includegraphics[width=0.5 \textwidth]{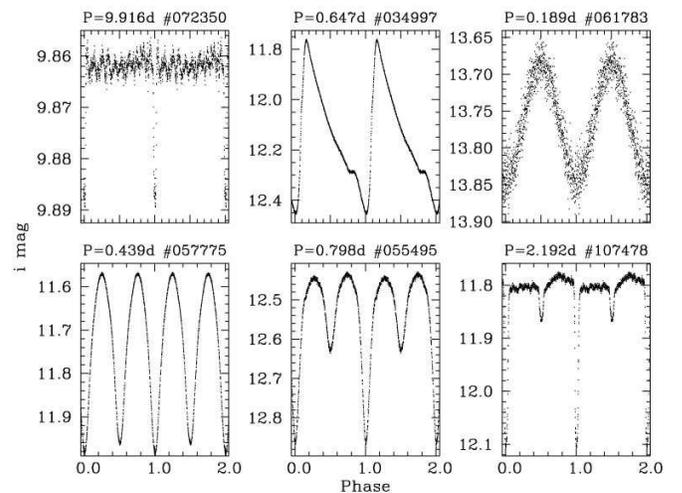}
\caption{Phased light curves of six variable stars. The periods and CSTAR IDs
  are given in the figure. Top row, from left to right: transit variable; RR
  Lyrae; $\delta$~Scuti. Bottom row, from left to right: eclipsing binaries of
  contact (EC), semi-detached (ESD), and detached (ED)
  configurations.\label{fig:lcphas}}
\end{center}
\end{figure}

The final sample of 157 variable stars is presented in Table
\ref{tb:variables}. Column 1 gives the CSTAR ID; column 2 gives the ID from the
Guide Star Catalog, version 2.3.2 (GSC2.3); columns 3 and 4 give the right
ascension and declination from GSC2.3; column 5 gives the mean $i$-band
magnitude; column 6 gives the main period, and column 7 gives a tentative
classification of the variable type, when possible. Fig.~\ref{fig:lcunph} shows
the time series of three bright stars (one constant and two
variables). Fig.~\ref{fig:lcphas} shows folded light curves of a representative
subset of the periodic variables, while Figs.~\ref{fig:lctimea}-f show the
time-series light curves for another representative subset of the periodic
variables. Lastly, Figs.~\ref{fig:lclpa}-b show two examples of variables with
long-term trends but no periodicity in our observing window.

28 CSTAR variables appear in previous variable-star catalogs: the All-Sky
Automated Survey \citep[ASAS,][]{Pojmanski2005} and the General Catalog of
Variable Stars \citep[GCVS,][]{Samus2009-GCVS}. The variables in common are
listed in Table~\ref{tb:prev_vars}. The periods we derived are in good agreement
with previous determinations except for CSTAR\#032007, \#052891 and \#136863.
The first two variables have CSTAR periods that are $\sim 2\times$ the ASAS
values, while the last one exhibits long-term variability in the CSTAR data which does
not match the ASAS period.

We were unable to recover 14 previously-known variables: 12 are
saturated and 2 lie very close to saturated stars and were masked in
our master frame. In the magnitude range $9.18 < i < 14.15$ that we
have in common with GCVS and ASAS, we found $5\times$ more variable
stars.

\ \par

We classified approximately half of the variables into one of five
types: binaries, $\delta$~Scuti, $\gamma$~Doradus, RR Lyrae, and
objects with long-term variability which may be periodic beyond our
observing window. One variable exhibits a light curve consistent with
an exoplanet transit. Table~\ref{tb:stats} contains crude statistics
for the different types.

\subsection{Search for multiple periods via Fourier Decomposition
 \label{sec:multper}}

We searched for multiple periods in the time-series photometry of the variables
using the Period04 program \citep{Lenz2005}. We started by identifying the
frequency ($f_1$) that displayed the highest $S/N$ peak in the periodogram.
Next we prewhitened the time series (i.e., subtracted off the most significant
frequency) and searched for the next highest peak in the frequency spectrum. We
repeated the process until all peaks with $S/N>4$ were identified. For example,
consider the variable CSTAR\#061353, shown in Fig.~\ref{fig:multp}. There are
three significant peaks in the periodogram, with $f_i=44.2879$, 44.1690, and
42.1209 cycles d$^{-1}$) and signal-to-noise ratios of 15.8, 15.7, and 15.0,
respectively. The top left inset shows the phased light curve for
$f_1=44.2879$~cycles d$^{-1}$ with 1000 phase bins. Given the
amplitude-frequency plot and the phased light curve, the dominant period of 32
minutes is likely real.

\ \par

The Fourier decomposition results of all the periodic variables are listed in
Table~\ref{tb:fourier}. We list in columns 2, 3, and 4 the frequency, amplitude
(in mmag), and signal-to-noise ratio of each peak. Column 5 gives the relations
between the primary and other frequencies. We do not list 4 transit-like
variables with $P>6$~d(072350, 081845, 097230, 131919) for which we only have a
period based on the BLS method.

\subsection{Types of variables found by CSTAR}

\subsubsection{Eclipsing binaries}

Eclipsing binaries can be classified into three broad categories
\citep{Paczynski2006a}: contact (EC), semi-detached (ESD) and detached
(ED). Typical light curves are shown in the three bottom panels of
Fig.~\ref{fig:lcphas} and in Figs.~15a, b and c.

\ \par

There are 10 binaries among the 27 variables in common with ASAS (CSTAR\#001707,
020526, 022489 036162, 038255, 038663, 055495, 057775, 083768, 133742). Their
Fourier spectra show that the frequencies found are often integral or
half-integral multiples of each other. We identified an additional 31 variables
exhibiting similar Fourier properties and light curves.

\ \par

\ \par

\ \par

\subsubsection{$\delta$ Scuti Stars}

$\delta$ Scuti variables are late A- and early F-type stars situated in the
instability strip on or above the main sequence in the HR Diagram. Their
typical pulsation periods are found to be in the interval of 0.02d to 0.25d \citep{Breger2000}. The frequencies of 9 candidates are in the range 4 to 30
cycles d$^{-1}$ such as star 061783 in Table \ref{tb:fourier}.

\vfill

\newcounter{subfigure}
\setcounter{subfigure}{1}
\renewcommand{\thefigure}{\arabic{figure}\alph{subfigure}}
\begin{figure}[h]%lc time series plots
\begin{center}
\includegraphics[width=0.45\textwidth]{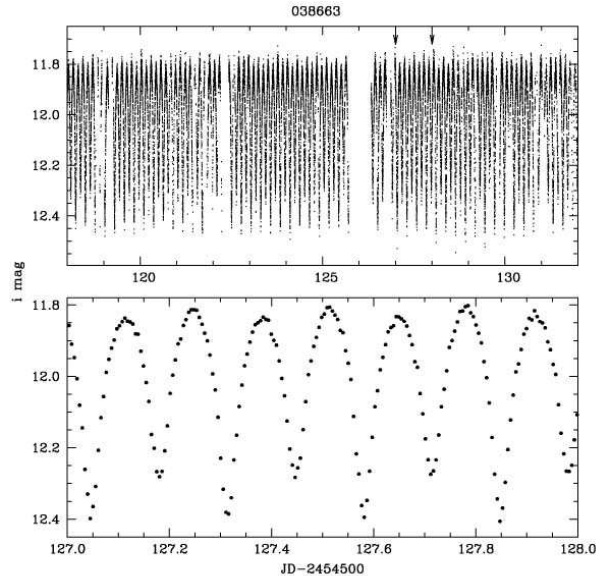}
\caption{Time-series light curve of a contact binary. Only a small fraction of
  the complete CSTAR data is shown. The top panel shows the light curve sampled
  at 20s intervals. The bottom panel shows a portion of the top light curve
  (bounded by the arrows) binned into 450s intervals. \label{fig:lctimea}}
\end{center}
\end{figure}

\addtocounter{figure}{-1}
\addtocounter{subfigure}{1}
\begin{figure}[b]
\begin{center}
\includegraphics[width=0.45\textwidth]{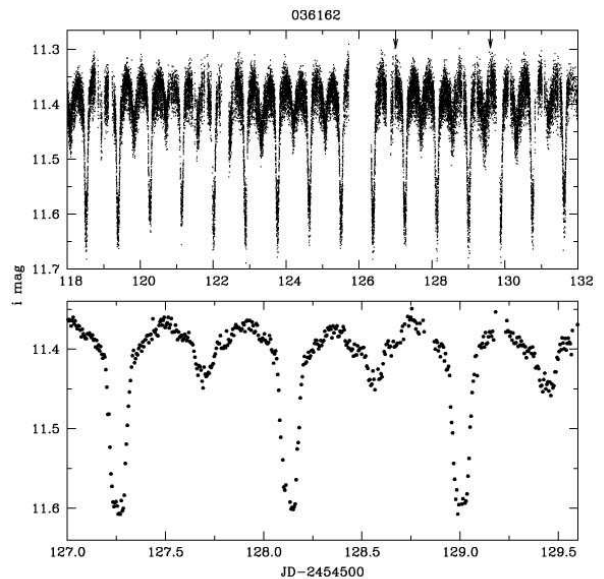}
\caption{Same as \ref{fig:lctimea}, but for a semi-detached
  binary.\label{fig:lctimeb}}
\end{center}
\end{figure}

\addtocounter{table}{1}
\LongTables
\begin{deluxetable*}{llllrrcrr}
\tablenum{3}
\tablewidth{0pt}
\tablecaption{Variable stars\label{tb:variables}}
\tablehead{\multicolumn{2}{c}{ID} & \colhead{R.A.} & \colhead{Dec.} & \colhead{$i$} & \multicolumn{2}{c}{Period} & \colhead{$T_0^3$} & \colhead{Type$^{4}$}\\
\colhead{CSTAR} & \colhead{GSC} & \multicolumn{2}{c}{(GSC$^{1}$)} & \colhead{(mag)} & \colhead{(d)} & \colhead{Src$^{2}$} & \colhead{(d)} & \colhead{}}
\startdata
000572  &  S3YM000353  & 11:11:19.60  & $-$87:18:00.9  & 13.88  &   0.154415 &  LS& 55.3409 &        DSCT \\
001707  &  S74D000321  & 12:32:42.91  & $-$87:26:22.9  & 11.07  &   0.338528 &  LS& 56.5058 &     EC, [A] \\
003125  &  S3YM000469  & 10:43:46.63  & $-$87:25:10.1  &  9.68  &   3.586394 &  LS&   \nd   &     Binary? \\
003697  &  S742000061  & 12:08:11.93  & $-$87:35:39.9  & 11.55  &  26.065850 &  LS&   \nd   &             \\
003850  &  S742000043  & 12:34:25.12  & $-$87:34:37.7  & 10.14  &  14.928409 & P04&   \nd   &         [A] \\
004463  &  S3YM000518  & 10:40:16.05  & $-$87:29:29.8  & 11.16  &   0.434167 &  LS& 55.5901 &     Binary? \\
005954  &  S742000078  & 12:39:58.23  & $-$87:41:37.2  & 11.68  &  74.844696 & P04&   \nd   &             \\
008426  &  S3YM000629  & 10:14:50.45  & $-$87:36:23.8  & 11.10  &   8.768682 &  LS&   \nd   &             \\
009171  &  S3YM000662  & 10:12:54.85  & $-$87:38:22.9  & 13.95  &   0.591609 &  LS& 55.9779 &    RRL, [A] \\
009952  &  S742000182  & 12:43:30.67  & $-$87:53:30.9  & 11.39  &  23.675364 & P04&   \nd   &         [A] \\
011616  &  S3Y9000240  & 10:56:28.86  & $-$87:55:21.0  & 11.97  &  30.935810 &  LS&   \nd   &             \\
011709  &  S742000246  & 12:41:44.27  & $-$87:58:28.5  & 11.40  &   2.950398 &  LS& 58.0466 &          EC \\
011796  &  S742000286  & 12:21:35.82  & $-$88:00:14.5  & 12.20  &   1.893822 &  LS& 56.5289 &         ESD \\
013140  &  S3Y9000236  & 10:25:53.99  & $-$87:53:40.8  &  9.81  &  20.532616 & P04&   \nd   &             \\
013255  &  S74F000377  & 14:54:21.37  & $-$87:21:05.0  &  9.84  &       \nd  & \nd&   \nd   &      LT, [A]\\
013432  &  S3YN000438  & 09:10:17.59  & $-$87:24:06.0  & 11.46  &  74.811104 & P04&   \nd   &             \\
014111  &  S74F000634  & 14:29:01.63  & $-$87:38:16.2  & 13.68  &   0.174075 &  LS& 55.4082 &          EC \\
014368  &  S3YM000753  & 10:03:10.79  & $-$87:51:06.2  & 10.77  &       \nd  & \nd&   \nd   &          LT \\
014495  &  S3Y9000339  & 10:49:00.65  & $-$88:02:17.2  & 12.24  &  18.711236 &  LS&   \nd   &             \\
016836  &  S742030458  & 12:49:16.22  & $-$88:11:17.6  & 13.70  &   0.176211 &  LS& 55.3351 &          EC \\
018708  &  S3YN000609  & 09:02:20.82  & $-$87:37:41.0  & 12.38  &   1.627682 &  LS& 56.1166 &     Binary? \\
020436  &  S3YN000517  & 08:40:33.28  & $-$87:28:38.6  & 12.55  &  55.208965 & P04&   \nd   &             \\
020526  &  S742000504  & 13:23:49.26  & $-$88:16:04.3  & 12.41  &   2.510798 &  LS& 56.6054 &    ESD, [A] \\
022489  &  S3Y9000527  & 10:01:21.80  & $-$88:13:30.8  & 11.88  &   0.652255 &  LS& 55.8390 &     EC, [A] \\
023614  &  S742000638  & 12:09:34.38  & $-$88:29:59.0  & 11.49  &       \nd  & \nd&   \nd   &          LT \\
023885  &  S742000650  & 12:01:19.22  & $-$88:30:46.1  & 11.17  &   1.737463 &  LS&   \nd   &             \\
025440  &  S742000656  & 12:12:56.28  & $-$88:34:11.6  & 10.78  &   9.823067 &  LS&   \nd   &             \\
025846  &  S742000671  & 14:25:39.12  & $-$88:14:54.1  & 13.76  &   9.781566 &  LS&   \nd   &             \\
026640  &  S3YB000429  & 11:17:00.40  & $-$88:35:36.4  &  9.19  &  49.365707 & P04&   \nd   &         [A] \\
026730  &  S3YB000202  & 09:58:36.02  & $-$88:23:59.9  & 12.89  &   2.065619 &  LS& 56.9905 &          ED \\
027082  &  S743000019  & 13:19:17.39  & $-$88:33:00.7  & 11.22  &   6.632957 &  LS&   \nd   &             \\
027575  &  S743000159  & 14:31:54.23  & $-$88:17:39.3  & 10.60  &   6.157666 &  LS&   \nd   &             \\
028221  &  S3YB000436  & 11:04:06.48  & $-$88:38:02.5  & 13.56  &   4.843837 &  LS&   \nd   &             \\
029379  &  S74E000029  & 15:57:05.68  & $-$87:30:05.4  & 11.95  &   3.102692 &  LS&   \nd   &CW-FO/EC, [A]\\
030353  &  S3YA000615  & 08:39:40.45  & $-$88:03:19.4  & 12.92  &   1.046833 &  LS& 56.3738 &             \\
032007  &  S3YB000225  & 09:29:39.14  & $-$88:30:06.4  & 11.77  &   0.621936 &  LS& 56.0565 &RRL, [A], [G]\\
032544  &  S3Y8000312  & 11:48:09.42  & $-$88:49:52.6  & 10.39  &  52.022440 &  LS& 89.0185 &             \\
034389  &  S3YB006313  & 09:49:58.18  & $-$88:41:27.2  & 14.04  &   1.735374 &  LS&   \nd   &             \\
034669  &  S743000094  & 13:50:03.38  & $-$88:46:13.2  & 10.06  &  14.610884 &  LS&   \nd   &             \\
034724  &  S3YB000482  & 10:01:18.94  & $-$88:44:36.8  & 10.07  &  42.966682 &  LS&   \nd   &DCEP-FU, [A] \\
034997  &  S743000311  & 14:29:04.38  & $-$88:38:43.7  & 12.08  &   0.646586 &  LS& 55.6868 &    RRL, [A] \\
035468  &  S3YA000613  & 08:08:46.28  & $-$88:00:02.0  & 13.81  &   0.822906 & BLS& 56.1514 &         ED? \\
036162  &  S3YB000243  & 09:03:59.29  & $-$88:33:07.6  & 11.40  &   0.873470 &  LS& 55.6252 &    ESD, [A] \\
036508  &  S74E000310  & 16:32:13.29  & $-$87:16:17.0  & 13.02  &  37.533190 &  LS&   \nd   &             \\
036526  &  S743000153  & 15:35:01.14  & $-$88:16:11.9  & 12.30  &       \nd  & \nd&   \nd   &          LT \\
036939  &  S743000200  & 15:25:11.74  & $-$88:23:14.9  & 12.82  &   7.384502 &  LS&   \nd   &             \\
037016  &  S743000186  & 15:28:41.36  & $-$88:21:44.0  & 13.60  &   0.157944 &  LS& 55.3402 &        DSCT \\
037271  &  S740000037  & 12:42:14.25  & $-$88:59:05.0  & 12.55  &  13.123393 &  LS&   \nd   &             \\
038255  &  S743000115  & 13:53:18.49  & $-$88:54:14.6  & 12.81  &   0.266903 &  LS& 55.5067 &     EC, [A] \\
038580  &  S3Y8000346  & 10:50:11.96  & $-$88:59:54.1  & 11.76  &   6.967593 &  LS&   \nd   &             \\
038663  &  S3YB000253  & 08:46:12.64  & $-$88:33:42.9  & 11.96  &   0.267127 &  LS& 55.4568 &     EC, [A] \\
039541  &  S740000060  & 12:36:24.67  & $-$89:04:02.7  & 10.97  &  13.745355 &  LS&   \nd   &             \\
040351  &  S3YA000660  & 08:02:26.52  & $-$88:14:26.1  & 13.33  &   0.285578 &  LS& 55.4676 &        DSCT \\
042081  &  S743000391  & 14:51:21.48  & $-$88:51:53.5  & 12.34  &   6.281005 &  LS&   \nd   &             \\
042266  &  S3YL000384  & 07:16:52.61  & $-$87:28:56.4  & 13.66  &   0.383158 &  LS& 57.3796 &          EC \\
043618  &  S3YB009826  & 08:40:28.89  & $-$88:47:00.4  & 13.69  &  12.998266 &  LS& 65.4350 &         ESD \\
043885  &  S3YB000275  & 08:17:16.96  & $-$88:37:29.5  & 13.90  &   9.439374 &  LS&   \nd   &             \\
044751  &  S740000101  & 13:01:29.04  & $-$89:13:47.0  & 13.18  &   6.495565 &  LS&   \nd   &             \\
047176  &  S3YA000181  & 07:12:14.54  & $-$87:51:19.7  & 12.79  &   2.626902 &  LS&   \nd   &     Binary? \\
048452  &  S3YB000382  & 08:21:18.73  & $-$88:55:48.2  & 13.40  &   9.975068 &  LS&   \nd   &             \\
048615  &  S743000407  & 15:44:38.97  & $-$88:54:18.9  & 13.44  &   2.919245 &  LS&   \nd   &             \\
050375  &  S3YB013754  & 07:39:17.11  & $-$88:40:41.7  & 13.45  &   1.440770 &  LS& 56.6100 &             \\
050773  &  S74E000469  & 17:10:42.44  & $-$87:30:09.3  & 10.55  &   3.265063 &  LS&   \nd   &     Binary? \\
052891  &  S741000155  & 17:13:18.69  & $-$87:42:28.6  &  9.50  &  53.957802 & P04&   \nd   &MISC/SR, [A] \\
053446  &  S3Y8000220  & 08:05:03.24  & $-$89:07:54.3  & 12.95  &   1.071097 &  LS& 55.9342 &     Binary? \\
053570  &  S743000460  & 16:44:10.40  & $-$88:38:19.9  & 11.02  &  19.238876 &  LS&   \nd   &             \\
053783  &  S3Y8000272  & 09:44:14.75  & $-$89:28:17.6  & 10.48  &  16.620420 &  LS&   \nd   &     Binary? \\
055495  &  S3Y8000078  & 07:43:54.49  & $-$89:07:37.3  & 12.51  &   0.797670 &  LS& 55.7331 &    ESD, [A] \\
055854  &  S3Y8000109  & 07:54:37.65  & $-$89:15:40.9  &  9.76  &  58.004639 & P04&   \nd   &             \\
057247  &  S741000463  & 17:18:35.08  & $-$88:12:47.1  & 13.45  &  42.032624 &  LS&   \nd   &             \\
057344  &  S741000539  & 17:13:42.52  & $-$88:24:52.5  & 11.08  &  20.533037 & P04&   \nd   &             \\
057775  &  S3YA000492  & 06:40:47.15  & $-$88:15:21.3  & 11.70  &   0.438611 &  LS& 55.5992 &     EC, [A] \\
059811  &  S3YA000336  & 06:28:42.76  & $-$88:02:41.7  & 12.36  &   7.259171 & BLS& 62.4178 &          ED \\
061353  &  S740000342  & 17:15:45.51  & $-$89:00:42.8  & 10.78  &   0.022581 & P04& 55.2804 &             \\
061658  &  S741000489  & 17:36:45.98  & $-$88:14:10.5  & 11.31  &   0.076167 & BLS& 55.3337 &        DSCT \\
061740  &  S740000322  & 17:23:25.05  & $-$88:53:37.3  &  9.97  &  25.220045 & P04&   \nd   &             \\
061783  &  S740000308  & 17:25:37.17  & $-$88:49:50.3  & 13.74  &   0.188991 &  LS& 55.4105 &        DSCT \\
062683  &  S3Y8000165  & 07:46:18.25  & $-$89:40:00.7  & 10.27  &  22.526073 & P04&   \nd   &             \\
062854  &  S3YA000596  & 06:23:15.53  & $-$88:33:35.4  & 12.68  &  34.733614 &  LS&   \nd   &             \\
063059  &  S3Y8000125  & 06:49:54.20  & $-$89:21:58.8  &  9.44  &  15.164152 & P04&   \nd   &             \\
063241  &  S3YA000025  & 06:12:09.29  & $-$87:27:15.6  & 13.81  &   0.131663 &  LS&   \nd   &        DSCT \\
064380  &  S3YA000248  & 06:10:28.12  & $-$87:53:32.9  & 14.14  &   2.309535 &  LS&   \nd   &             \\
064944  &  S741000460  & 17:51:13.16  & $-$88:09:48.8  & 10.63  &  12.420856 &  LS&   \nd   &             \\
065072  &  S740000301  & 17:47:26.35  & $-$88:46:07.9  & 11.87  &   0.620578 &  LS& 55.4131 &             \\
066196  &  S740000469  & 17:05:16.14  & $-$89:51:43.8  &  9.65  &  45.493835 & P04&   \nd   &             \\
066682  &  S3YA021353  & 06:03:05.44  & $-$88:29:29.3  & 13.43  &   1.272036 &  LS&   \nd   &             \\
066775  &  S741000378  & 17:59:00.73  & $-$88:01:32.9  & 11.76  &  27.296301 & P04&   \nd   &             \\
068276  &  SA9S000144  & 18:22:33.29  & $-$89:36:22.9  & 11.40  &   2.796789 &  LS& 55.2942 &     Binary? \\
068308  &  S0SG000353  & 05:50:34.20  & $-$89:06:46.2  & 12.93  &   0.798475 &  LS&   \nd   &             \\
068493  &  S0SH000282  & 05:55:30.36  & $-$87:57:06.8  & 14.02  &   0.714096 & BLS&   \nd   &             \\
068908  &  SA9U000383  & 18:08:15.09  & $-$88:18:02.9  & 10.83  &   2.645011 &  LS&   \nd   &             \\
069430  &  S0SH026837  & 05:48:54.94  & $-$88:30:16.6  & 14.02  &  13.380021 &  LS&   \nd   &             \\
070680  &  SA9S000413  & 22:05:02.55  & $-$89:52:06.7  & 13.01  &   1.987754 &  LS& 55.9380 &         ESD \\
070941  &  S0SH000215  & 05:47:08.05  & $-$87:51:00.2  & 10.22  &   0.606466 &  LS& 55.7721 &        GDOR \\
071571  &  S0SH000333  & 05:43:19.93  & $-$88:04:04.3  & 10.57  &       \nd  & \nd&   \nd   &     LT, [A] \\
072350  &  SA9V000050  & 18:30:57.87  & $-$88:43:17.5  &  9.86  &   9.916101 & BLS& 63.6590 &     transit \\
072730  &  SA9U000438  & 18:29:03.93  & $-$88:32:31.9  & 13.26  &   0.573063 &  LS& 55.7163 &    RRL, [A] \\
073028  &  S0SH000497  & 05:30:59.30  & $-$88:30:04.7  & 13.52  &   9.675598 &  LS&   \nd   &             \\
073846  &  SA9U000442  & 18:35:37.14  & $-$88:33:45.9  & 11.97  &  12.501163 &  LS&   \nd   &             \\
076135  &  SA9S000115  & 19:53:21.15  & $-$89:22:46.5  & 10.49  &       \nd  & \nd&   \nd   &          LT \\
076723  &  SA9V000063  & 19:00:22.60  & $-$88:47:53.7  & 10.84  &  51.559681 & P04&   \nd   &             \\
077171  &  SA9S015549  & 23:03:19.23  & $-$89:39:40.4  & 10.10  &   5.681234 &  LS&   \nd   &             \\
077190  &  S0SH000448  & 05:15:49.62  & $-$88:17:51.5  & 10.20  &  49.365650 & P04&   \nd   &             \\
077508  &  SA9S000404  & 22:53:06.46  & $-$89:38:40.9  & 14.04  &   0.133311 &  LS& 55.4247 &        DSCT \\
077594  &  SA9V000067  & 19:08:12.87  & $-$88:49:18.5  & 10.06  &   1.677415 &  LS& 56.7822 &             \\
078549  &  S0SH000321  & 05:15:49.03  & $-$88:04:23.5  & 13.10  &   5.393342 &  LS& 65.7215 &         EC? \\
078773  &  SA9V000073  & 19:17:53.08  & $-$88:51:11.2  & 11.81  &   0.186027 &  LS& 55.2921 &          EC \\
079397  &  SA9S000068  & 19:48:57.10  & $-$89:07:14.3  & 10.86  &   4.851947 &  LS&   \nd   &             \\
080934  &  SA9U000331  & 18:58:35.51  & $-$88:12:55.2  & 11.74  &  64.449600 & P04&   \nd   &             \\
081428  &  S0SG000131  & 00:46:05.01  & $-$89:31:34.7  & 10.38  &  10.643167 & P04&   \nd   &             \\
081563  &  SA9V000069  & 19:33:28.81  & $-$88:48:44.8  & 12.77  &  23.342061 &  LS&   \nd   &             \\
081723  &  SA9U000071  & 18:48:23.75  & $-$87:43:16.4  & 11.48  &   0.841544 &  LS& 56.0909 &        GDOR \\
081749  &   $\dots$    & 19:11:53.61  & $-$88:27:14.9  & 10.30  &  70.308655 & P04&   \nd   &             \\
081845  &  SA9V000074  & 19:38:27.80  & $-$88:50:55.9  & 12.89  &   6.767879 & BLS& 56.0302 &         ED? \\
082180  &  S0SG000129  & 00:29:58.19  & $-$89:30:20.3  & 11.24  &   0.152807 &  LS& 55.3621 &         EC? \\
082489  &  S0SG000205  & 03:11:22.58  & $-$89:15:13.6  & 14.08  &   0.172334 &  LS& 55.6132 &          EC \\
083110  &  S0SG000226  & 02:11:02.80  & $-$89:22:50.3  & 10.27  &  40.003201 & P04&   \nd   &             \\
083768  &  S0SU000304  & 05:13:29.62  & $-$87:19:42.6  & 12.04  &   0.384076 &  LS& 55.5749 &     EC, [A] \\
084344  &  SA9S000371  & 23:01:30.03  & $-$89:25:01.5  & 13.94  &   1.466449 &  LS&   \nd   &             \\
085531  &  SA9U000338  & 19:20:09.67  & $-$88:14:03.9  & 13.66  &   6.367062 &  LS&   \nd   &             \\
085719  &  S0SG000178  & 03:08:19.12  & $-$89:06:32.2  & 12.19  &  16.573029 & P04&   \nd   &             \\
086480  &  S0SJ000306  & 03:44:10.88  & $-$88:52:09.5  & 11.19  &  23.017941 &  LS&   \nd   &             \\
087084  &  SA9S000168  & 20:57:31.47  & $-$89:03:50.3  & 12.39  &   1.857031 & BLS& 56.6568 &         ED? \\
087501  &  S0SG000092  & 01:23:01.27  & $-$89:17:09.4  & 13.54  &   0.193486 &  LS& 55.3480 &         EC? \\
087548  &  S0SH000485  & 04:20:11.85  & $-$88:25:03.5  & 12.62  &   0.197741 &  LS& 55.3917 &          EC \\
088142  &  S0SG000093  & 00:52:40.76  & $-$89:17:32.4  & 13.75  &   0.292953 &  LS& 55.3454 &          EC \\
089391  &  SA9U000387  & 19:43:40.77  & $-$88:19:20.7  & 13.48  &   0.136882 &  LS&   \nd   &             \\
093873  &  SA9S000300  & 22:17:44.44  & $-$89:01:38.1  &  9.42  &  34.120377 & P04&   \nd   &             \\
094793  &  S0SU000407  & 04:36:31.02  & $-$87:28:03.3  & 10.96  &   0.616082 &  LS&   \nd   &             \\
096404  &  SAA5000420  & 19:43:05.07  & $-$87:46:52.4  & 13.93  &   0.581793 &  LS& 55.9001 &    RRL, [A] \\
097230  &  SA9U000295  & 20:02:18.84  & $-$88:02:50.0  & 12.06  &  19.253283 & BLS& 66.7927 &         ED? \\
097790  &  SA9V000415  & 22:23:40.80  & $-$88:53:42.9  &  9.71  &   0.521650 &  LS& 55.4856 &        GDOR \\
098092  &  S0SH000022  & 03:54:41.13  & $-$88:02:50.6  & 11.63  &  89.237907 & P04&   \nd   &         [A] \\
098719  &  SAA5000417  & 19:50:26.13  & $-$87:44:50.7  & 12.80  &   0.208220 &  LS& 55.3070 &          EC \\
099529  &  S0SG000018  & 00:31:15.83  & $-$88:55:17.9  & 10.79  &   4.833741 &  LS&   \nd   &             \\
106019  &  S0SU000388  & 04:02:44.72  & $-$87:22:26.5  & 10.60  &  32.131606 &  LS&   \nd   &             \\
106769  &  SA9V000337  & 22:04:27.19  & $-$88:29:34.7  & 13.66  &  24.117975 &  LS&   \nd   &             \\
107478  &  SAA5000503  & 20:28:30.07  & $-$87:46:16.5  & 11.80  &   2.192167 &  LS& 55.7692 &      ED/ESD \\
110801  &  S0SI000269  & 02:42:27.87  & $-$88:04:22.5  &  9.48  &  35.695164 & P04&   \nd   &         [A] \\
111298  &  S0SI000438  & 02:12:56.05  & $-$88:13:52.5  & 10.01  &  11.922923 &  LS&   \nd   &             \\
113453  &  S0SI000393  & 00:45:26.78  & $-$88:25:21.3  & 11.57  &  34.526609 &  LS&   \nd   &             \\
114506  &  S0SV000507  & 02:55:24.43  & $-$87:48:18.6  & 12.09  &   3.852912 & BLS& 57.4873 &             \\
116410  &  S0SV000502  & 02:47:02.44  & $-$87:47:06.9  & 11.60  &   3.693423 &  LS&   \nd   &             \\
116471  &  S0SV000340  & 03:22:43.36  & $-$87:23:34.4  & 13.44  &   0.263255 &  LS&   \nd   &       DSCT? \\
117654  &  S0ST000503  & 02:18:11.67  & $-$87:56:11.2  & 10.54  &  21.683977 & P04&   \nd   &             \\
118705  &  S0SI033411  & 00:53:26.68  & $-$88:12:41.4  & 13.37  &   8.461614 &  LS&   \nd   &             \\
119488  &  S0SI000329  & 00:08:43.43  & $-$88:13:48.4  & 10.07  &  80.006401 & P04&   \nd   &             \\
120188  &  S0SI000289  & 00:55:45.92  & $-$88:09:11.0  & 11.16  &  10.711880 &  LS&   \nd   &             \\
121369  &  SA9T000310  & 23:15:35.46  & $-$88:07:33.8  & 10.77  &  21.092596 & P04&   \nd   &             \\
121389  &  S0SV000324  & 03:02:08.67  & $-$87:22:53.3  & 13.73  &  44.773727 &  LS&   \nd   &             \\
123934  &  SA9T000282  & 23:52:30.36  & $-$88:03:49.5  & 13.91  &   0.122021 &  LS&   \nd   &       DSCT? \\
124666  &  SAA6000034  & 21:47:16.29  & $-$87:39:06.6  & 12.75  &   0.458038 &  LS& 55.5358 &    RRL, [A] \\
127850  &  S0SI014387  & 00:00:52.46  & $-$87:54:27.3  & 11.79  &       \nd  & \nd&   \nd   &     LT, [G] \\
128178  &  S0SI000097  & 01:08:04.79  & $-$87:47:59.1  & 13.87  &  15.846209 &  LS&   \nd   &             \\
131919  &  S0SI000101  & 00:01:16.84  & $-$87:44:02.9  & 11.97  &   9.448529 & BLS& 57.2732 &          ED \\
133742  &  SAA6000366  & 22:37:07.30  & $-$87:28:49.9  & 11.58  &   0.848378 &  LS& 55.9822 &     EC, [A] \\
136863  &  SAA6000283  & 23:01:23.88  & $-$87:22:19.9  &  9.50  &       \nd  & \nd&   \nd   &     LT, [A]
\enddata
\tablecomments{[1]: if the star is not in GSC2.3, the position is
  based on the CSTAR master image. [2] Source of period: [LS], Lomb-Scargle
  method; [BLS], box fitting algorithm; [P04], Fourier decomposition with
  Period04 program. [3] Epoch of primary eclipse or minimum light (when
  applicable), in JD-2454500. [4] Type: [A], ASAS variable; [G], GCVS variable;
  CW-FO, W Virginis variable, first-overtone pulsator; DCEP-FU, $\delta$~Cephei
  variable, fundamental-mode pulsator; DSCT, $\delta$~Scuti variable; EC,
  contact binary; ESD, semi-detached binary; ED, detached binary; GDOR,
  $\gamma$~Doradus variable; LT, long-term trend; MISC/SR,
  Miscellaneous/semi-regular variable; RRL, RR~Lyrae variable.}
\end{deluxetable*}
\addtocounter{table}{3}

\addtocounter{table}{-2}
\begin{deluxetable}{llrr}
\tablewidth{0pt}
\tablenum{4}
\tablecaption{Variables in common between CSTAR and other catalogs\label{tb:prev_vars}}
\tablehead{\multicolumn{2}{c}{ID} & \multicolumn{2}{c}{$P$ (d)}\\
\colhead{CSTAR} & \colhead {Other} & \colhead{CSTAR} & \colhead{Other}}
\startdata
001707 & [A]123244-8726.4 &  0.338528 &  0.338519 \\
003850 & [A]123423-8734.6 & 14.928409 & 18.63354  \\
009171 & [A]101257-8738.4 &  0.591609 &  0.59161  \\
009952 & [A]124330-8753.5 & 23.675364 & 23.71541  \\
013255 & [A]145422-8721.1 & LT        &174.6807   \\
020526 & [A]132341-8816.1 &  2.510798 &  2.51046  \\
022489 & [A]100123-8813.5 &  0.652255 &  0.65226  \\
026640 & [A]111701-8835.6 & 49.365707 & 51.8      \\
029379 & [A]155703-8730.1 &  3.102692 &  3.115622 \\
032007 & [A]092907-8829.7 &  0.621936 &  0.38355  \\
034724 & [A]100112-8844.6 & 42.966682 & 43.5      \\
034997 & [A]142904-8838.7 &  0.646586 &  0.64655  \\
036162 & [A]090354-8833.1 &  0.873470 &  0.87376  \\
038255 & [A]135318-8854.2 &  0.266903 &  0.266899 \\
038663 & [A]084613-8833.7 &  0.267127 &  0.267128 \\
052891 & [A]171319-8742.5 & 53.957802 & 29.036005 \\
055495 & [A]074400-8907.7 &  0.797670 &  0.79801  \\
057775 & [A]064047-8815.4 &  0.438611 &  0.43863  \\
071571 & [A]054317-8804.1 &  LT       &426        \\
072730 & [A]182912-8832.6 &  0.573063 &  0.573034 \\
083768 & [A]051332-8719.7 &  0.384076 &  0.38408  \\
096404 & [A]194301-8746.9 &  0.581793 &  0.58171  \\
098092 & [A]035443-8802.8 & 89.237907 & 70.7      \\
110801 & [A]024230-8804.4 & 35.695164 & 47.6      \\
124666 & [A]214719-8739.1 &  0.458038 &  0.45803  \\
127850 & [G] ST Oct       &  LT       &  LT       \\
133742 & [A]223703-8728.8 &  0.848378 &  0.84838  \\
136863 & [A]230125-8722.3 &  LT       & 37.31343
\enddata
\tablecomments{[A]: ASAS; [G]: GCVS; LT=long-term variability.}
\end{deluxetable}

\addtocounter{table}{1}
\begin{deluxetable}{lrr}
\tablewidth{0pt}
\tablenum{5}
\tablecaption{Distribution of variable star types \label{tb:stats}}
\tablehead{\colhead{Variable Type} & \colhead{N} & \colhead{\%}}
\startdata
Binary       & 41 & 26.1 \\
$\delta$~Sct &  9 &  5.7 \\
Long-term    &  8 &  5.1 \\
RR Lyr       &  6 &  3.8 \\
$\gamma$~Dor &  3 &  1.9 \\
Transit      &  1 &  0.6 \\
Unclassified & 89 & 56.7
\enddata
\end{deluxetable}
\vspace{-12pt}

\addtocounter{figure}{-1}
\addtocounter{subfigure}{1}
\begin{figure}[t]
\begin{center}
\includegraphics[width=0.45\textwidth]{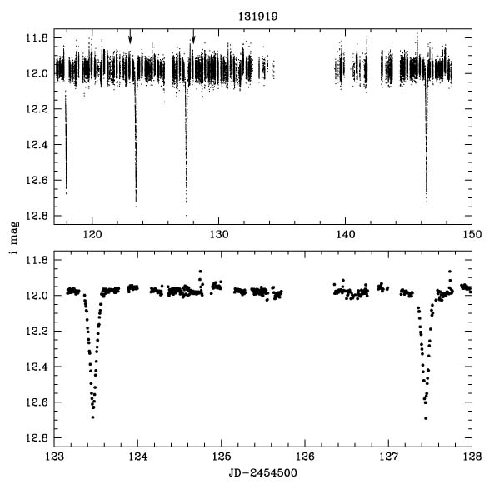}
\caption{Same as \ref{fig:lctimea}, but for a detached
  binary.\label{fig:lctimec}}
\end{center}
\end{figure}
\addtocounter{figure}{-1}
\addtocounter{subfigure}{1}

\begin{figure}[b]
\begin{center}
\includegraphics[width=0.45\textwidth]{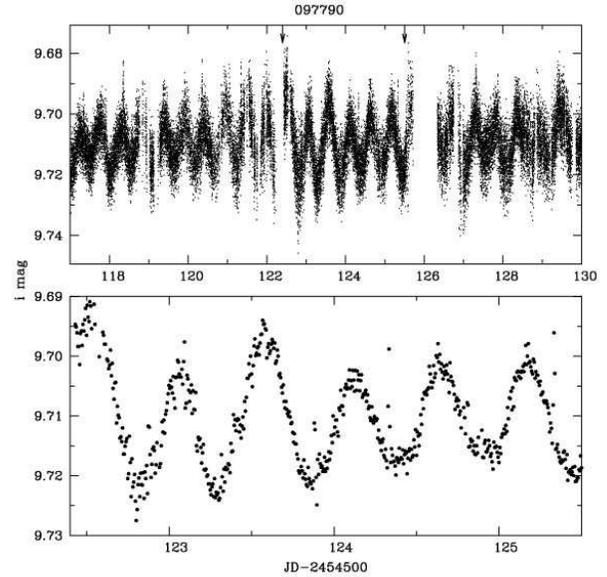}
\caption{Same as \ref{fig:lctimea}, but for a variable of $\gamma$~Dor
  type.\label{fig:lctimed}}
\end{center}
\end{figure}

\vfill\clearpage

\addtocounter{figure}{-1}
\addtocounter{subfigure}{1}
\begin{figure}[t]
\begin{center}
\includegraphics[width=0.45\textwidth]{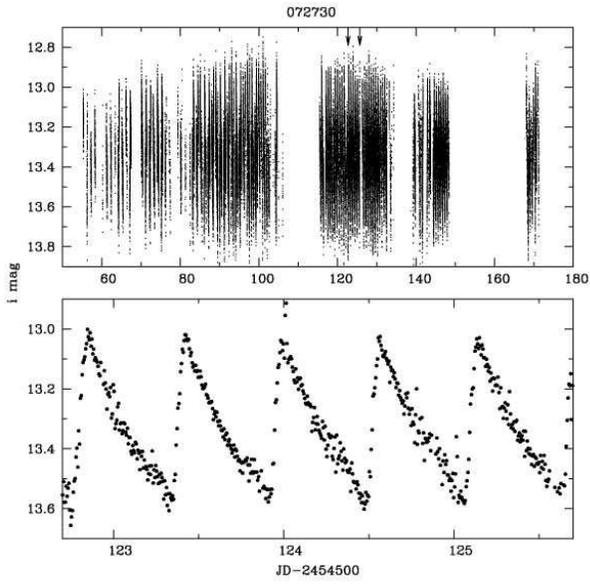}
\caption{Same as \ref{fig:lctimea}, but for a variable of RR~Lyr type with
  Blazhko effect.\label{fig:lctimee}}
\end{center}
\end{figure}

\vfill

\setcounter{subfigure}{1}
\begin{figure}[b]
\includegraphics[width=0.45\textwidth]{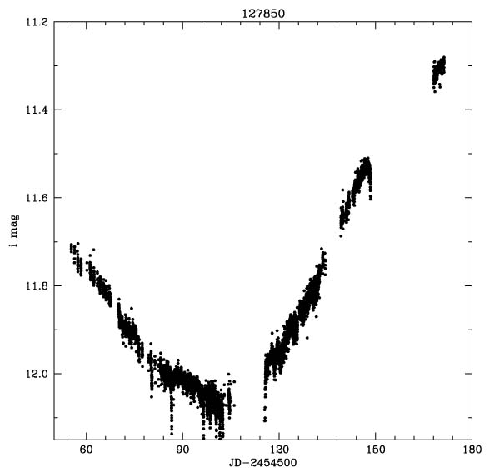}
\caption{Complete CSTAR light curve for a long-term variable, binned at 450s intervals.\label{fig:lclpa}}
\end{figure}

\begin{figure}[t]
\begin{center}
\addtocounter{figure}{-2}
\addtocounter{subfigure}{5}
\includegraphics[width=0.45\textwidth]{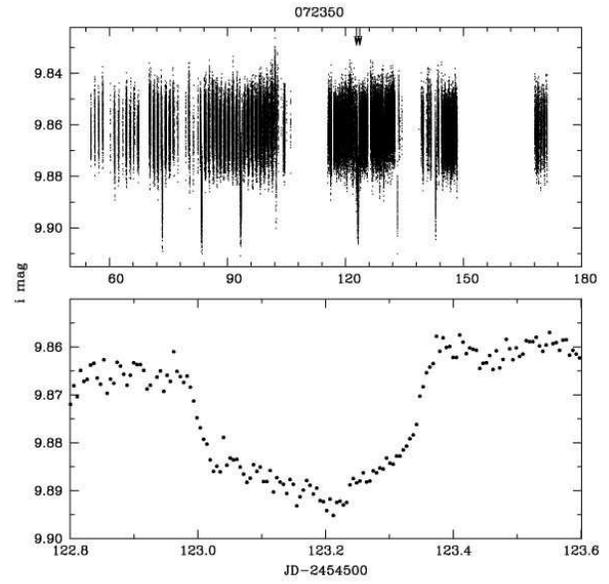}
\caption{Same as \ref{fig:lctimea}, but for a possible transiting
  exoplanet.\label{fig:lctimef}}
\end{center}
\end{figure}

\vspace*{2.05in}

\addtocounter{subfigure}{-4}
\begin{figure}[b]
\includegraphics[width=0.45\textwidth]{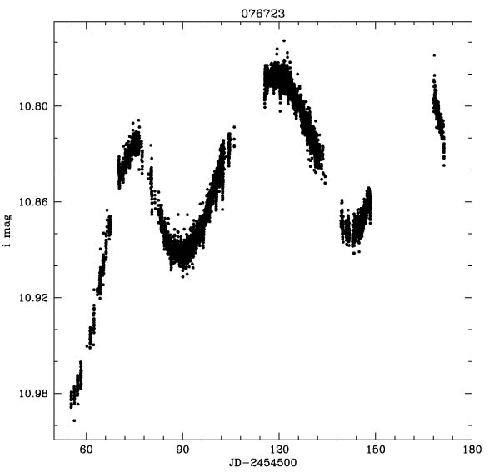}
\caption{Complete CSTAR light curve for a long-term variable, binned at 450s intervals.\label{fig:lclpb}}
\end{figure}
\renewcommand{\thefigure}{\arabic{figure}}

\clearpage

\begin{figure}[t]
\begin{center}
\includegraphics[width=0.45\textwidth]{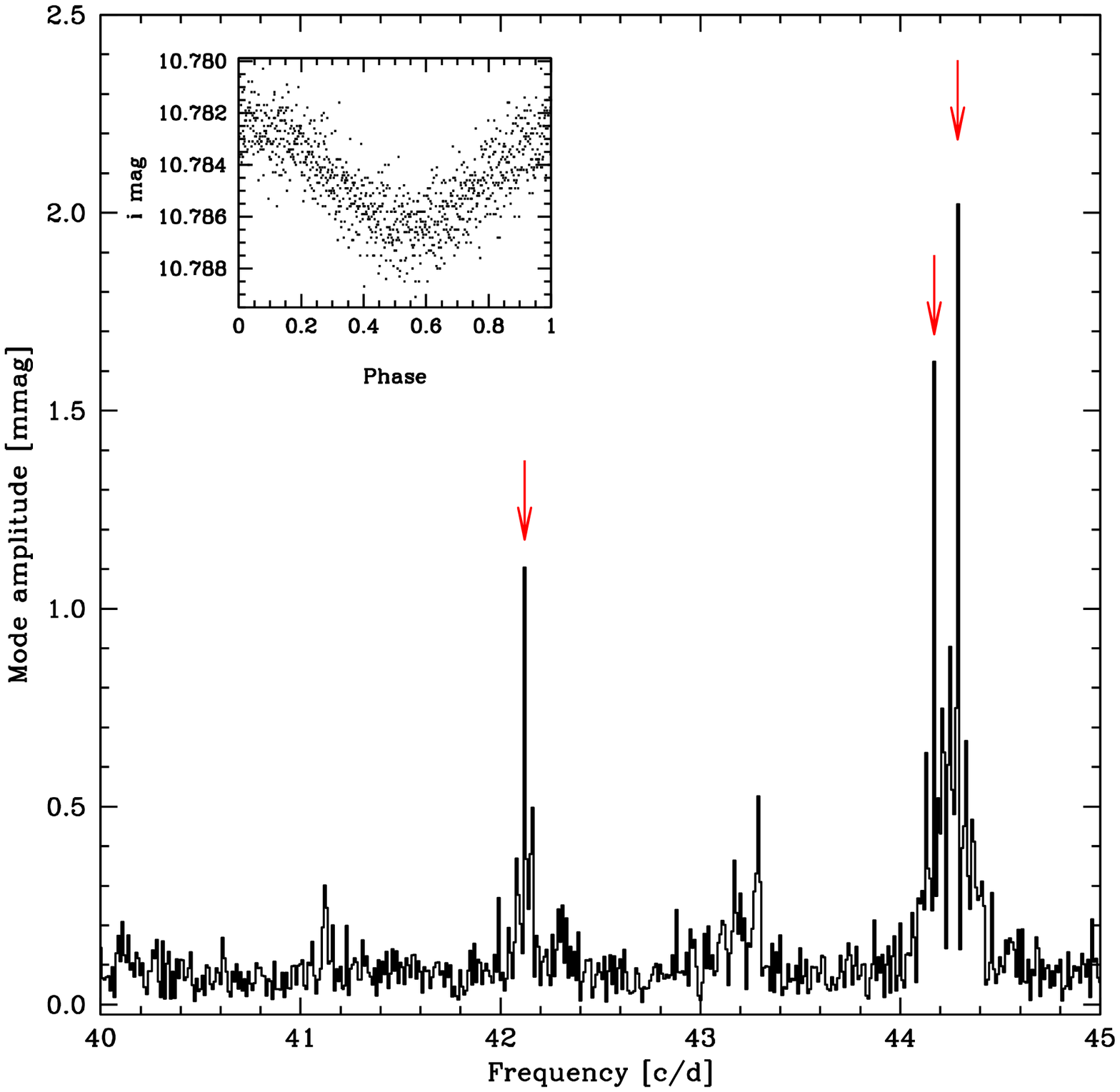}
\caption{Fourier spectrum of variable candidate ID = 061353 derived using the
  Period04 program. There are three significant peaks, at f$_i$ = 44.288,
  44.169, and 42.121 cycles d$^{-1}$. The inset on the top-left corner shows
  the light curve phased using f$_i$=44.288 cycles d$^{-1}$.  See
  Table~\ref{tb:fourier} for details. \label{fig:multp}}
\end{center}
\end{figure}

\subsubsection{$\gamma$ Doradus Stars}

$\gamma$ Doradus variables constitute a class of multi-periodic pulsating
variables that have a strong asteroseismic potential since they are located in
a region of the HR Diagram where $g$-mode pulsations, $p$-mode pulsations and
solar-like variability may exist in a single star \citep{Kaye1999}. They are of
spectral type A or F and are located on or just above the main
sequence. Observationally, they exhibit multiple periods between 0.3 and 3 d,
with typical photometric amplitudes of 5 to 50 mmag \citep{Cuypers2009a}. Based
on the Fourier decomposition, three candidates are likely $\gamma$ Dor
stars. Star 070941 has three frequencies of 1.65, 1.60, and 1.83 cycles
d$^{-1}$. Its spectral type is F0 from the Tycho-2 Spectral Type Catalog 
\citep{Wright2003}. It is located at RA = 05:47:08.050, DEC:$-$87:51:00.17
(J2000). To our knowledge there are no published spectra of the other two
candidates (081723, and 097790 shown in Fig.~\ref{fig:lctimed}).

\subsubsection{RR Lyrae Stars}

RR Lyrae are low-mass stars undergoing helium core fusion with spectral types
ranging from A2 to F6 \citep{Smith1995}. They exhibit periods from $\sim0.2$ to
$\sim1$~day and photometric amplitudes in the optical bands of 0.3 to 2
mag. Most RR Lyrae stars pulsate in the radial fundamental mode (RRab
stars). RRc stars pulsate radially in the first overtone. RRd stars, which are
much rarer, pulsate in both modes simultaneously. There are six RRAB stars 
in our sample (009171, 032007, 034997, 072730, 096404, 124666), all of which were
previously identified by ASAS. Additionally, CSTAR\#032007 is listed in the GCVS as
W Oct.

Some RR Lyrae stars show modulations in their period, pulsation amplitude,
light curve shape, and radial velocity curves. These modulations are known as
the ``Blazhko effect'' \citep{Blazhko1907,Buchler2011}. Previous surveys have
found different percentages of RR Lyrae exhibiting the Blazhko effect: 10-30\%
of RRab variables in \citet{Alcock2003}, and 50\% in \citet{Jurcsik2009a}. The
Fourier spectrum of a Blazhko-effect RRLyrae is characterized by triplets,
quintuplets and high-order multiplets around the pulsation frequency
components.

Based on the period analysis in Table \ref{tb:fourier} and light curve
shapes we found 4 out of 6 RR~Lyr stars in our sample exhibit the Blazhko
effect (034997, 072730, 124666, 032007). One such star is shown in
Fig.~\ref{fig:lctimee}.

\subsubsection{Other Types of Variables}

Variable CSTAR\#072350 displays a transit-like light curve with a period of
9.916d and a depth of 17~mmag, shown in the top left panel of
Fig.~\ref{fig:lcphas} and Fig.~\ref{fig:lctimef}. The primary is a bright F5
star with $T_{\rm eff} \sim 6400$~K from the Tycho-2 Spectral Type
Catalog \citep{Wright2003}. If confirmed, this would be an interesting addition
to the known extrasolar planets given the relatively long orbital period and
temperature of the primary.

There are 8 variable candidates with long-term trends, which may have periods
longer than the observing window of 128~d. Some are probably Mira stars, such
as CSTAR\#127850 which shows an amplitude variation of 1.4 mag. Its light curve
is plotted in Fig.~\ref{fig:lclpa}.

89 variable candidates have no classification in column 6 of Table
\ref{tb:variables}; half of these were previously unknown. Follow-up
photometric and spectroscopic observations will be required to determine their
nature. One of these variables, CSTAR\#065072, has been spectroscopically
identified as a M3.5 dwarf in the solar neighborhood by \citet{Riaz2006AJ-MDwarfCatalog}.

\subsection{Blending}

The large pixel scale of the camera and the relatively high stellar density of
the field makes blending a problem. We define a blend as two or more stars
located within 30 arcseconds ($\sim$ 2 CSTAR pixels) of each other. Inspection
of DSS images reveals that $\sim 1/3$ of the candidate variable stars are
blended with one neighboring star while $\sim 6$\% are blended with two or more
neighboring stars. Follow-up photometry with a finer pixel scale will be
required to ensure these blends are not responsible for the observed
variability. Future telescopes planned for Dome A will yield a considerably
finer pixel scale which will mitigate this problem
\citep[AST3,][]{Cui2008SPIE-AST3}. Additionally, we are implementing a
difference-imaging photometry pipeline that will enable a more robust detection
of variables in crowded environments and will deliver higher photometric
precision.

\section{Summary}
  
CSTAR has a large field of view (23 square degrees) and the ability to provide
uninterrupted observations for the entire duration of the Antarctic winter
night. We have obtained high-quality time-series photometry for
10,000 stars with $i<14.5$~mag and detected 157 variables, $5\times$ more
than previous surveys of same area of the sky in the same magnitude range.

Our photometry indicates that Dome A is a viable and excellent observing site.
During the 2008 Antarctic winter season, 96\% of the images obtained at a solar
elevation angle below $-10^\circ$ were useful for scientific purposes.  The
median sky background in the $i$ band was 19.6 mag/$\sq \arcsec$ and the median
extinction due to clouds was below 0.1 mag.

\acknowledgements

Lingzhi Wang acknowledges financial support by the China Scholarship
Council and the National Natural Science Foundation of China under the
Distinguished Young Scholar Grant 10825313 and Grant 11073005, by the
Ministry of Science and Technology National Basic Science Program
(Project 973) under grant number 2007CB815401, and by the Excellent
Doctoral Dissertation of Beijing Normal University Engagement Fund.

\ \par

\ \par

\ \par

Lucas Macri and Lifan Wang acknowledge support by the Department
of Physics \& Astronomy at Texas A\&M University through faculty startup funds
and the Mitchell-Munnerlyn-Heep Chair for tenure-track faculty. 

This work was supported by the Chinese PANDA International Polar Year project,
NSFC-CAS joint key program through grant number 10778706, CAS main direction
program through grant number KJCX2-YW-T08. The authors deeply appreciate the
great efforts made by the 24-27th Dome A expedition teams who provided
invaluable assistance to the astronomers that set up and maintained the CSTAR
telescope and the PLATO system. PLATO was supported by the Australian Research
Council and the Australian Antarctic Division. Iridium communications were
provided by the US National Science Foundation and the US Antarctic
Program. 
\bibliography{paper}{}
\bibliographystyle{apj}

\clearpage

\begin{appendix}
\renewcommand{\thetable}{A\arabic{table}}
\setcounter{table}{0}
\LongTables
\begin{deluxetable}{lrrrr}
\tablewidth{0pt}
\tablenum{1}
\tablecaption{Fourier Analysis of Variable Star Candidates\label{tb:fourier}}
\tablehead{\colhead{CSTAR} & \colhead{Freq.} & \colhead{Amp.} &
\colhead{$S/N$} & \colhead{Notes}\\
\colhead{ID} & \colhead{[c/d]} & \colhead{[mmag]} &  &}
\startdata
000572 &    6.475955 &   60.74 & 11.04 &       \\   &   16.964241 &   11.56 &  4.26 &       \\ \hline
001707 &    5.907794 &  259.90 & 19.84 &    f1 \\   &   11.815589 &   94.63 & 19.71 &   2f1 \\   &   17.723383 &   43.02 & 20.49 &   3f1 \\   &   23.631178 &   16.74 & 18.22 &   4f1 \\   &    2.953241 &   13.88 &  7.30 & 0.5f1 \\   & 29.539444 &    7.67 & 15.23 &   5f1 \\ \hline
003125 &    0.278487 &   22.41 &  8.53 &    f1 \\   &    0.139674 &   16.74 &  8.38 & 0.5f1 \\   &    5.016641 &    5.56 &  4.27 &       \\ \hline
003697 &    0.037071 &   33.69 & 13.28 &       \\ \hline
004463 &    2.303088 &    7.66 &  4.74 &    f1 \\   &    5.013947 &    9.09 &  5.98 &       \\   &    4.604452 &    6.45 &  5.49 &   2f1 \\   &    6.905816 &    6.21 &  6.45 &   3f1 \\   &    9.206318 &    4.97 &  6.43 &   4f1 \\   &   11.510546 &    3.98 &  7.64 &   5f1 \\   &   13.812341 &    2.76 &  6.52 &   6f1 \\   &   13.032144 &    2.50 &  6.10 &       \\   &   16.113611 &    2.18 &  5.81 &   7f1 \\   &   26.071184 &    1.27 &  4.27 &       \\   &   34.092800 &    1.02 &  4.33 &       \\ \hline
005954 &    0.013361 &   37.42 & 16.30 &       \\   &    0.051720 &   10.34 &  7.87 &       \\ \hline
008426 &    0.113784 &   10.93 &  7.71 &       \\ \hline
009171 &    1.690393 &  270.20 & 10.48 &    f1 \\   &    3.380355 &  139.87 & 10.87 &   2f1 \\   &    5.070748 &   90.51 & 10.35 &   3f1 \\   &    6.761141 &   67.58 & 10.71 &   4f1 \\   &    8.451533 &   42.15 &  9.40 &   5f1 \\   & 10.141927 &   20.77 &  5.92 &   6f1 \\ \hline
009952 &    0.042238 &   30.29 & 11.46 &       \\ \hline
011616 &    0.031463 &   17.31 & 10.59 &       \\ \hline
011709 &    0.339196 &   23.99 & 13.45 &    f1 \\   &    0.675806 &    7.24 &  7.16 &   2f1 \\ \hline
011796 &    1.057236 &   67.40 & 11.70 &    f1 \\   &    2.114041 &   40.08 & 11.25 &   2f1 \\   &    3.171277 &   30.60 & 11.68 &   3f1 \\   &    0.528834 &   30.16 & 10.52 & 0.5f1 \\   &    1.585208 &   22.13 & 11.80 & 1.5f1 \\   &    4.228082 &   18.44 & 11.99 &   4f1 \\   &    2.642013 &   18.03 & 13.69 & 2.5f1 \\   &    3.699680 &   12.52 & 13.01 & 3.5f1 \\   &    5.285318 &    9.32 & 14.17 &   5f1 \\   &    4.756054 &    6.72 & 12.99 & 4.5f1 \\ \hline
013140 &    0.048703 &   19.44 & 11.78 &       \\   &    0.021550 &    7.78 &  7.46 &       \\ \hline
013432 &    0.013367 &   26.49 &  9.41 &       \\   &    0.040102 &   18.95 &  7.52 &       \\   &    0.073305 &   11.12 &  4.96 &       \\   &    5.012740 &    8.36 &  6.03 & \\ \hline
014111 &    5.744743 &  172.54 & 17.85 &    f1 \\   &   11.489056 &   27.26 & 10.16 &   2f1 \\   &    2.868061 &   15.18 &  4.39 & 0.5f1 \\ \hline
014495 &    0.053444 &   25.15 & 14.02 &       \\   &    0.037066 &    6.09 &  6.47 &       \\ \hline
016836 &    5.675009 &  122.96 & 23.33 &    f1 \\   &    0.102148 &   20.23 &  6.06 &       \\   &   11.350019 &   17.23 & 13.80 &   2f1 \\   &    0.070684 &   16.05 &  5.12 &       \\   &   17.026321 &    5.29 &  5.39 &   3f1 \\ \hline
018708 &    1.228764 &   15.06 &  7.68 &    f1 \\   &    2.457959 &    9.67 &  6.90 &   2f1 \\   &    3.687154 &    7.94 &  6.86 &   3f1 \\   &    0.614382 &    7.48 &  5.86 & 0.5f1 \\   &   28.365950 &    4.86 & 10.15 &       \\   &    4.916780 &    3.88 &  5.57 &   4f1 \\   &   24.912863 &    2.18 &  4.95 &       \\   &    1.057168 &    4.68 &  5.16 &       \\   &   27.730873 &    2.04 &  4.85 & \\   &    1.173146 &    3.23 &  3.42 & \\   &   11.060168 &    1.82 &  4.01 &   9f1 \\   &    9.833129 &    2.36 &  5.12 &   8f1 \\   &   25.758816 &    1.47 &  3.92 & \\   &   26.705179 &    1.34 &  3.75 & \\   &    4.298518 &    2.36 &  3.38 & 3.5f1 \\   &    8.602209 &    1.64 &  3.73 &   7f1 \\   &   24.896910 &    1.47 &  3.54 & \\   &   23.724195 &    1.65 &  4.52 & \\ \hline
020436 &    0.018113 &  115.20 & 14.09 &       \\   &    0.027600 &   23.34 &  9.13 &       \\ \hline
020526 &    0.796911 &  133.14 & 13.22 &    f1 \\   &    1.593822 &   50.02 & 10.83 &   2f1 \\   &    1.195151 &   34.28 &  9.21 & 1.5f1 \\   &    0.398671 &   30.44 &  9.87 & 0.5f1 \\ \hline
022489 &    3.066545 &  107.03 & 25.84 &    f1 \\   &    6.132659 &    7.88 & 37.04 &   2f1 \\   &    1.533057 &    7.63 &  9.71 & 0.5f1 \\   &    4.599602 &    4.20 & 10.45 & 1.5f1 \\   &    3.059649 &    3.89 &  6.19 &       \\   &    9.199203 &    2.17 & 10.60 &   3f1 \\ \hline
023885 &    0.575379 &    3.02 &  5.34 &       \\ \hline
025440 &    0.101715 &    3.72 &  7.21 &       \\   &    0.210326 &    3.16 &  7.30 &       \\ \hline
025846 &    0.101716 &   27.26 &  8.53 &       \\ \hline
026640 &    0.020257 &   31.03 & 12.39 &       \\   &    0.009482 &   37.92 & 13.69 &       \\   &    0.036635 &   10.34 &  8.42 &       \\   &    0.053013 &    6.14 &  6.53 &       \\ \hline
026730 &    0.967161 &   11.96 &  8.56 &    f1 \\   &    2.904499 &    8.66 &  7.85 &   3f1 \\   &    4.840114 &    7.44 &  8.12 &   5f1 \\   &    1.936046 &    8.39 &  6.32 &   2f1 \\   &    3.872522 &    8.02 &  7.83 &   4f1 \\   &    6.776160 &    5.37 &  7.93 &   7f1 \\   &    2.419195 &    6.48 &  6.11 & 2.5f1 \\   &    5.808568 &    6.17 &  7.74 &   6f1 \\   &    0.483149 &    5.71 &  5.35 & 0.5f1 \\   &    7.744183 &    5.59 &  8.85 &   8f1 \\   &    4.356534 &    4.98 &  7.14 & 4.5f1 \\ \hline
027082 &    0.150418 &    2.85 &  6.26 &       \\ \hline
027575 &    0.162054 &    5.04 &  6.59 &       \\   &    0.074562 &    4.74 &  6.66 &       \\   &    0.179294 &    2.99 &  5.65 &       \\ \hline
028221 &    0.206017 &   71.45 & 13.86 &       \\   &    0.026722 &    9.72 &  4.83 &       \\   &    0.102146 &   10.48 &  4.45 &       \\ \hline
029379 &    0.322387 &   23.21 &  7.89 &       \\ \hline
030353 &    0.955090 &   46.94 & 15.46 &       \\   &    1.024911 &    5.72 &  4.49 &       \\   &   28.078613 &    1.68 &  4.01 &       \\ \hline
032007 &    1.608054 &   44.58 & 16.98 &    f1 \\   &    3.215677 &   15.04 & 11.48 &   2f1 \\   &    1.645551 &   12.70 &  7.01 &    f3 \\   &    4.823300 &    6.18 &  9.95 &   3f1 \\   &    1.571419 &    5.78 &  4.32 &    f2 \\   &    3.253174 &    6.09 &  5.92 &   2f3 \\   &    4.861228 &    4.86 &  7.27 &   3f3 \\   &    6.469282 &    4.11 &  9.72 &   4f1 \\ \hline
032544 &    0.036635 &   17.74 & 17.24 &       \\   &    0.044393 &    6.70 &  8.61 &       \\   &    0.057754 &    4.91 &  8.21 &       \\   &    0.021119 &    6.40 & 15.52 & \\ \hline
034389 &    0.576245 &   24.15 &  7.39 &       \\ \hline \hline
034669 &    0.068097 &   19.74 & 13.55 &       \\   &    0.061201 &    6.87 &  9.08 &       \\   &    0.010344 &    3.81 &  7.02 &       \\ \hline
034724 &    0.024136 &   65.58 & 14.39 &    f1 \\   &    0.048272 &   15.11 &  8.82 &   2f1 \\   &    0.030601 &    9.54 &  6.71 &       \\   &    0.056892 &    6.71 &  5.54 &       \\ \hline
034997 &    1.546844 &  227.25 & 19.62 &    f1 \\   &    3.093256 &  122.24 & 19.93 &   2f1 \\   &    4.640100 &   79.59 & 20.65 &   3f1 \\   &    6.186513 &   52.77 & 22.37 &   4f1 \\   &    7.733357 &   26.78 & 20.52 &   5f1 \\   &    9.279770 &   16.35 & 19.04 &   6f1 \\   &   10.826613 &   10.94 & 17.50 &   7f1 \\   &   12.373026 &    5.82 & 14.94 &   8f1 \\   &    1.539948 &    7.04 &  6.56 &       \\ \hline
035468 &    2.430840 &   15.96 &  6.73 &    f1 \\   &    7.292090 &    9.30 &  6.18 &   3f1 \\   &    4.861250 &    8.85 &  6.21 &   2f1 \\   &    9.721638 &    8.65 &  6.93 &   4f1 \\ \hline
036162 &    2.289026 &   63.01 & 16.56 &    f1 \\   &    1.144729 &   34.01 & 12.84 & 0.5f1 \\   &    4.577621 &   29.43 & 14.61 &   2f1 \\   &    3.433323 &   25.90 & 17.49 & 1.5f1 \\   &    6.866647 &   18.35 & 15.33 &   3f1 \\   &    5.722350 &   17.01 & 20.09 & 2.5f1 \\   &    8.010513 &    9.12 & 15.74 & 3.5f1 \\   &    9.155673 &    7.67 & 17.89 &   4f1 \\ \hline
036508 &    0.026729 &   85.37 & 10.56 &       \\   &    3.958093 &   18.96 &  4.52 &       \\ \hline
036939 &    0.135333 &    8.50 &  8.41 &       \\ \hline
037016 &    6.331342 &   29.84 & 18.14 &       \\ \hline
037271 &   0.006896  &   17.65 &  7.69 &       \\   &    0.076286 &   15.95 &  5.71 &       \\   &    0.065080 &   11.36  & 4.93 &       \\ \hline
038255 &    7.493285 &  239.91 & 27.40 &    f1 \\   &   14.986570 &   70.71 & 27.40 &   2f1 \\   &   22.480286 &   21.95 & 23.99 &   3f1 \\   &    3.746643 &   17.78 & 14.17 & 0.5f1 \\   &    0.935260 &    7.69 &  4.73 &       \\   &    0.988703 &    7.97 &  5.10 &       \\   &   37.466766 &    4.13 & 10.37 &   5f1 \\   &   26.226521 &    3.66 &  9.09 & 3.5f1 \\   &    6.017149 &    3.44 &  5.70 &       \\   &    7.502790 &    3.24 &  5.50 &       \\   &   44.960915 &    3.00 &  7.94 &   6f1 \\   &   33.718830 &    2.80 &  7.88 & 4.5f1 \\   &   14.995214 &    2.72 &  5.64 &       \\ \hline
038580 &    0.143522 &    7.60 &  7.78 &       \\   &    0.174122 &    6.88 &  7.28 &       \\ \hline
038663 &    7.487258 &  227.63 & 27.71 &    f1 \\   &   14.974086 &   64.44 & 26.97 &   2f1 \\   &    3.743629 &   28.80 & 24.60 & 0.5f1 \\   &   22.461344 &   25.91 & 26.46 &   3f1 \\   &   11.230456 &   14.91 & 24.69 & 1.5f1 \\   &   18.718145 &    9.14 & 23.23 & 2.5f1 \\   &   29.948694 &    6.86 & 21.60 &   4f1 \\   &    7.479931 &    5.35 & 12.48 &       \\   &   26.204634 &    3.04 & 14.97 & 3.5f1 \\   &    3.753542 &    2.74 &  8.33 &       \\   &   14.965915 &    1.71 &  8.66 &       \\   &    7.493742 &    3.52 &  6.06 &       \\   &    7.502361 &    2.30 &  9.23 &       \\   &   37.435745 &    1.55 &  8.87 &   5f1 \\   &    7.472192 &    1.43 &  4.68 &       \\   &    9.022915 &    0.96 &  4.68 &       \\   &   13.035054 &    0.93 &  4.72 &       \\   &   14.981862 &    1.19 &  4.99 &       \\   &   33.692978 &    0.87 &  5.42 & 4.5f1 \\   &    7.463141 &    1.08 &  3.79 &      \\ \hline
039541 &    0.072838 &   13.94 & 14.34 &       \\   &    0.044392 &    5.18 &  6.17 &       \\   &    0.017240 &    4.42 &  7.19 &       \\   &    0.090509 &    3.55 &  6.86 &       \\ \hline
040351 &    3.501848 &   25.48 & 17.69 &       \\   &    3.477713 &   19.49 & 16.44 &       \\ \hline
042081 &    0.159469 &    6.45 &  9.15 &       \\ \hline
042266 &    5.219686 &  183.05 & 17.38 &    f1 \\   &   10.439812 &   49.96 &  5.83 &   2f1 \\   &   15.658175 &   19.28 &  8.27 &   3f1 \\   &   25.066116 &    9.26 &  4.50 & 5f1-1 \\ \hline
043618 &    0.153436 &   92.61 & 13.78 &    f1 \\   &    0.308595 &   37.63 &  8.33 &   2f1 \\   &    0.461600 &   33.14 &  7.75 &   3f1 \\   &    0.615467 &   22.20 &  7.15 &   4f1 \\   &    0.077149 &   21.86 &  6.65 & 0.5f1 \\   &    0.768902 &   13.83 &  6.63 &   5f1 \\   &    0.229292 &   13.74 &  6.44 & 1.5f1 \\   &    0.380141 &   12.41 &  4.79 & 2.5f1 \\ \hline
043885 &    0.106025 &   50.37 & 12.65 &       \\ \hline
044751 &    0.153434 &   14.38 & 10.57 &       \\ \hline
047176 &    0.380418 &   40.79 & 11.63 &    f1 \\   &    0.762992 &   10.60 &  6.09 &   2f1 \\ \hline
048452 &    0.100422 &   19.16 & 10.39 &       \\   &    0.173261 &    7.87 &  5.50 &       \\ \hline
048615 &    0.343072 &   10.86 &  7.56 &       \\ \hline
050375 &    0.693901 &   27.68 & 11.98 &       \\   &    0.035342 &    8.66 &  5.69 &       \\   &    0.103008 &    5.76 &  4.33 &       \\ \hline
050773 &    0.306100 &   23.02 &  8.15 &    f1 \\   &    0.612632 &   10.47 &  4.75 &   2f1 \\ \hline
052891 &    0.018533 &   64.30 & 20.88 &       \\   &    0.071545 &    5.00 &  5.81 &       \\   &    0.029308 &    5.97 &  5.77 &       \\   &    0.057753 &    5.19 &  4.55 &       \\   &    0.208171 &    3.27 &  4.36 &       \\ \hline
053446 &    0.933536 &   27.67 & 11.48 &    f1 \\   &    1.871813 &   15.00 &  8.98 &   2f1 \\   &    0.103008 &    8.83 &  5.23 &       \\ \hline
053570 &    0.051288 &    5.50 &  9.98 &       \\   &    0.075855 &    2.84 &  6.28 &       \\ \hline
053783 &    0.060770 &    3.45 &  8.76 &    f1 \\   &    0.123265 &    2.26 &  6.27 &   2f1 \\ \hline
055495 &    2.506240 &  134.34 & 20.81 &    f1 \\   &    1.253336 &   61.92 & 17.65 & 0.5f1 \\   &    5.012481 &   55.40 & 19.32 &   2f1 \\   &    3.759576 &   34.08 & 21.97 & 1.5f1 \\   &    7.518721 &   23.69 & 18.48 &   3f1 \\   &    6.265816 &   17.99 & 21.71 & 2.5f1 \\   &    8.772057 &    6.93 & 14.34 & 3.5f1 \\   &   10.025393 &    5.24 & 13.85 &   4f1 \\   &    0.100853 &    2.99 &  4.55 &       \\   &   11.277453 &    2.04 &  6.68 & 4.5f1 \\   &   16.043404 &    2.00 &  6.15 & 6.4f1 \\   &   15.034443 &    1.82 &  5.98 &   6f1 \\   &    7.019646 &    1.40 &  4.26 & 2.8f1 \\   &   40.113522 &    1.32 &  4.66 &  16f1 \\ \hline
055854 &    0.017240 &    8.42 & 10.33 &       \\   &    0.059478 &   10.44 &  8.22 &       \\   &    0.026722 &    7.40 &  7.66 &       \\   &    0.070683 &    4.89 &  6.36 &       \\   &    0.040945 &    4.52 &  6.05 &       \\ \hline
057247 &    0.024136 &  103.94 & 15.85 &       \\ \hline
057344 &    0.048702 &   29.96 & 12.44 &       \\   &    0.036204 &   14.65 & 12.65 &       \\   &    0.010344 &   12.20 & 11.31 &       \\   &    0.018964 &    4.96 &  6.98 &       \\   &    0.089216 &    4.77 &  5.86 &       \\ \hline
057775 &    4.559935 &  186.04 & 27.32 &    f1 \\   &    9.119870 &   42.91 & 26.41 &   2f1 \\   &   13.679375 &   14.87 & 25.22 &   3f1 \\   &    6.839903 &    4.85 & 13.70 & 1.5f1 \\   &    2.278675 &    2.85 &  4.68 & 0.5f1 \\   &    4.553039 &    3.18 &  6.48 &    f2 \\   &   18.239328 &    1.98 &  9.93 &   4f1 \\   &   15.957636 &    1.21 &  6.81 & 3.5f1 \\   &    9.112992 &    1.43 &  5.14 &   2f2 \\   &   20.518433 &    1.03 &  6.06 & 4.5f1 \\ \hline
059811 &    0.551742 &   27.24 &  5.13 &    f1 \\   &    1.103485 &   26.13 &  5.69 &   2f1 \\   &    1.655227 &   22.77 &  5.53 &   3f1 \\   &    0.275871 &   27.04 &  5.52 & 0.5f1 \\   &    0.828044 &   24.20 &  6.17 & 1.5f1 \\   &    1.379356 &   22.14 &  6.38 & 2.5f1 \\   &    1.931529 &   20.21 &  6.33 & 3.5f1 \\   &    2.207400 &   19.30 &  6.47 &   4f1 \\   &    2.758711 &   16.57 &  5.97 &   5f1 \\   &    2.482840 &   16.22 &  5.91 & 4.5f1 \\   &    3.034582 &   14.31 &  5.75 & 5.5f1 \\   &    3.310885 &   13.31 &  5.69 &   6f1 \\   &    3.586325 &   10.94 &  5.31 & 6.5f1 \\   &    3.862627 &    9.52 &  4.89 &   7f1 \\ \hline
061353 &   44.287922 &    2.11 & 15.79 &       \\   &   44.168968 &    1.62 & 15.75 &       \\   &   42.120911 &    1.18 & 15.07 &       \\   &   37.660107 &    0.36 &  5.56 &       \\   &   39.444427 &    0.35 &  5.65 &       \\   &   42.308395 &    0.25 &  4.23 & \\   &   45.354671 &    0.23 &  3.74 & \\ \hline
061658 &   13.129424 &    4.07 & 19.30 &       \\   &   23.308718 &    2.03 & 11.76 &       \\   &   22.974697 &    1.24 &  8.36 &       \\   &   16.556738 &    1.07 &  8.14 &       \\   &   20.022375 &    0.95 &  7.00 &       \\   &   26.205441 &    0.92 &  7.64 & \\   &   23.396641 &    0.80 &  6.37 & \\   &   18.800934 &    0.73 &  6.03 & \\   &   11.388237 &    0.69 &  5.61 & \\   &   10.866733 &    0.69 &  5.81 & \\   &   12.723462 &    0.52 &  4.86 & \\   &   23.211313 &    0.52 &  4.49 & \\ \hline
061740 &    0.039651 &    8.28 & 11.18 &       \\   &    0.090940 &    3.72 &  8.01 &       \\   &    0.018533 &    7.11 &  7.18 &       \\   &    0.055167 &    5.25 &  5.06 &       \\ \hline
061783 &    5.291335 &   81.82 & 18.91 &       \\   &    8.022555 &   10.23 &  7.05 &       \\   &    7.014887 &    6.97 &  4.20 &       \\ \hline
062683 &    0.044393 &    5.83 &  9.62 &       \\   &    0.063357 &    7.04 &  8.23 &       \\   &    0.028015 &    6.08 &  6.70 &       \\   &    0.083182 &    4.41 &  7.10 &       \\ \hline
062854 &    0.029739 &    7.82 &  9.51 &       \\ \hline
063059 &    0.065945 &    6.78 &  8.60 &       \\   &    0.032757 &    6.23 &  7.10 &       \\   &    0.024568 &    5.12 &  5.12 &       \\   &    0.075427 &    3.93 &  5.37 &       \\   &    0.124994 &    2.48 &  5.03 &       \\ \hline
063241 &    7.595059 &   38.45 & 11.47 &       \\   &    5.013037 &   12.85 &  4.73 &       \\ \hline
064380 &    0.432298 &   20.95 &  5.31 &       \\   &    0.102579 &   15.94 &  4.47 &       \\ \hline
064944 &    0.080596 &   10.34 & 11.50 &       \\   &    0.104301 &    7.13 &  8.34 &       \\   &    0.054736 &    4.91 &  6.56 &       \\   &    0.032756 &    3.38 &  4.23 &       \\ \hline
065072 &    1.611488 &    8.86 & 11.01 &       \\   &    2.350212 &    7.61 & 11.21 &       \\   &    2.673889 &    4.78 &  8.84 &       \\   &    1.171443 &    4.16 &  7.17 &       \\ \hline
066196 &    0.021981 &    6.74 & 10.55 &       \\   &    0.074131 &    7.30 &  8.36 &       \\   &    0.061201 &    3.72 &  7.09 &       \\   &    0.112490 &    3.69 &  6.39 &       \\   &    0.096974 &    2.70 &  5.27 &       \\ \hline
066682 &    0.786142 &    9.78 &  5.42 &       \\   &    0.103009 &    8.70 &  4.99 &       \\ \hline
066775 &    0.036635 &   34.98 & 14.47 &       \\   &    0.010344 &   15.26 &  9.11 &       \\   &    0.019395 &   15.08 & 13.33 &       \\   &    0.056461 &    7.51 &  6.71 &       \\   &    0.027584 &   14.51 &  5.06 &       \\   &    0.086200 &    3.72 &  4.32 & \\   &    0.203862 &    2.77 &  4.60 & \\ \hline
068276 &    0.357294 &    9.73 & 12.18 &    f1 \\   &    0.713296 &    3.88 &  7.16 &   2f1 \\ \hline
068308 &    1.252474 &   14.79 &  8.00 &       \\ \hline
068493 &    0.074132 &   81.11 &  4.18 &       \\ \hline
068908 &    0.377984 &    4.51 &  8.38 &       \\   &    0.368502 &    1.74 &  4.37 &       \\   &    0.021981 &    1.45 &  4.26 &       \\   &    0.352124 &    1.58 &  3.87 &       \\ \hline
069430 &    0.074997 &   76.28 & 12.56 &       \\ \hline
070680 &    1.006812 &   51.96 & 14.67 &       \\   &    0.500389 &   23.21 &  9.49 &       \\   &    2.014054 &   21.48 & 12.01 &       \\   &    0.509009 &   13.26 &  9.84 &       \\   &    3.018711 &   11.81 & 11.17 &       \\   &    1.512803 &    8.11 &  7.83 & \\   &    2.517029 &    4.86 &  6.14 & \\   &    4.007421 &    4.91 &  6.04 & \\   &    1.502890 &    5.09 &  4.61 & \\   &    5.012939 &    3.52 &  5.48 & \\   &    0.057323 &    3.33 &  4.35 & \\ \hline
070941 &    1.648985 &    9.23 &  8.97 &       \\   &    1.600282 &    5.04 &  7.32 &       \\   &    1.829140 &    2.84 &  4.60 &       \\ \hline
071571 &    0.034480 &   38.60 & 13.56 &       \\   &    0.022843 &   30.81 & 10.19 &       \\   &    0.084906 &    4.31 &  5.38 &       \\ \hline
072730 &    1.745096 &  183.43 & 18.05 &    f1 \\   &    3.490192 &   95.66 & 18.02 &   2f1 \\   &    5.235289 &   61.37 & 18.30 &   3f1 \\   &    6.980816 &   37.60 & 17.47 &   4f1 \\   &    8.725912 &   19.39 & 13.96 &   5f1 \\   &    1.732597 &   12.05 &  7.19 &    f2 \\   &    1.757595 &   12.05 &  9.28 &    f3 \\   &   10.470577 &   10.51 & 10.45 &   6f1 \\   &    3.502260 &    8.41 &  7.46 &   2f3 \\   &    5.247356 &    7.54 &  7.53 &   3f3 \\   &    6.993315 &    7.05 &  6.57 &   4f3 \\   &   12.215673 &    5.91 &  7.29 &   7f1 \\   &    6.968748 &    6.42 &  6.79 &   4f2 \\   &    5.222359 &    5.44 &  6.35 &   3f2 \\   &    8.022099 &    5.15 &  5.52 &       \\   &   10.483938 &    3.87 &  5.14 &   6f3 \\   &    3.477694 &    4.03 &  5.09 &   2f2 \\   &   20.942879 &    3.62 &  4.73 &  12f1 \\   &   19.194765 &    3.57 &  4.90 &  11f1 \\   &   24.432209 &    3.56 &  4.75 &  14f1 \\   &   22.685820 &    3.42 &  4.93 &  13f1 \\   &    9.019851 &    3.61 &  4.78 & \\ \hline
073028 &    0.103439 &   15.11 &  5.60 &       \\ \hline
073846 &    0.080165 &    5.91 & 10.40 &       \\ \hline
076723 &    0.019395 &   38.94 & 12.92 &       \\   &    0.041807 &   36.84 &  7.73 &       \\   &    0.059909 &   21.68 &  7.91 &       \\   &    0.075856 &   18.13 &  7.96 &       \\   &    0.092664 &   12.76 &  9.34 &       \\   &    0.112059 &    7.73 &  9.01 & \\   &    0.122834 &    3.71 &  5.99 & \\   &    0.175847 &    2.32 &  5.27 & \\ \hline
077171 &    0.176277 &    2.61 &  6.32 &       \\ \hline
077190 &    0.020257 &    5.03 &  8.09 &       \\   &    0.048272 &    6.55 &  8.02 &       \\   &    0.080165 &    4.90 &  7.21 &       \\ \hline
077508 &    7.501159 &   54.76 & 15.81 &       \\ \hline
077594 &    0.596069 &    3.91 & 10.88 &       \\ \hline
078549 &    0.185328 &   60.49 & 15.03 &    f1 \\   &    0.093957 &    9.05 &  6.32 & 0.5f1 \\ \hline
078773 &    5.375379 &   30.51 & 26.14 &    f1 \\   &    2.687258 &    5.54 & 11.71 & 0.5f1 \\   &    0.013361 &    2.17 &  5.69 &       \\   &    2.677777 &    1.98 &  6.74 &       \\   &    2.697171 &    1.67 &  4.91 &       \\   &    2.012750 &    1.72 &  5.39 & \\   &    0.958966 &    1.37 &  3.96 & \\   &    5.368483 &    1.16 &  4.13 & \\   &    6.017132 &    0.89 &  4.70 & \\   &    7.022645 &    0.66 &  3.89 & \\   &    8.062206 &    0.64 &  4.03 & \\   &    0.971033 &    1.22 &  3.61 & \\   &   18.047947 &    0.64 &  4.48 & \\   &   46.353683 &    0.51 &  4.09 & \\ \hline
079397 &    0.206448 &    5.05 & 10.94 &       \\   &   37.988960 &    1.24 & 11.76 &       \\   &   32.580379 &    1.07 & 10.37 &       \\   &   35.173904 &    1.01 & 10.74 &       \\   &   32.170715 &    0.61 &  7.12 &       \\   &    8.019998 &    0.51 &  6.08 & \\   &   35.604038 &    0.46 &  5.46 & \\   &   39.746784 &    0.44 &  5.38 & \\ \hline
080934 &    0.015516 &   44.90 & 14.39 &       \\   &    0.034480 &   19.99 & 13.38 &       \\   &    0.024567 &   17.00 & 13.71 &       \\   &    0.052151 &    3.41 &  4.95 &       \\ \hline
081428 &    0.093957 &    3.26 &  6.14 &       \\   &    0.049996 &    2.68 &  5.04 &       \\ \hline
081563 &    0.042669 &    7.92 &  8.21 &       \\ \hline
081723 &    1.188292 &   20.55 &  9.77 &       \\   &    1.299493 &    6.84 &  4.18 &       \\   &    1.340869 &    5.42 &  3.58 &       \\ \hline
081749 &    0.014223 &    5.12 & 12.45 &       \\ \hline
082180 &    6.544276 &    4.83 & 18.95 &    f1 \\   &    0.027584 &    3.60 &  6.06 &       \\   &    0.158176 &    2.42 &  4.79 &       \\   &    0.074563 &    1.98 &  4.45 &       \\   &    3.011984 &    1.36 &  5.25 &       \\   &    4.012330 &    1.20 &  5.79 & \\   &   13.087427 &    1.08 &  7.51 &   2f1 \\ \hline
082489 &    5.802769 &  173.02 & 15.26 &    f1 \\   &    2.900522 &   34.11 &  5.80 & 0.5f1 \\   &   11.605107 &   33.53 &  7.75 &   2f1 \\   &   17.407015 &   15.47 &  4.96 &   3f1 \\ \hline
083110 &    0.024998 &    5.28 &  6.86 &       \\   &    0.117231 &    3.02 &  5.79 &       \\ \hline
083768 &    5.207372 &  185.73 & 14.02 &    f1 \\   &    5.013339 &   26.76 &  9.42 &       \\   &   10.414312 &   30.34 & 13.63 &   2f1 \\   &    7.022822 &   15.31 &  5.51 &       \\   &    2.606279 &   10.40 &  5.15 & 0.5f1 \\   &   20.829922 &    8.56 &  8.71 &   4f1 \\   &    2.010348 &    7.57 &  4.37 &       \\   &   26.036289 &    6.58 &  8.86 &   5f1 \\   &    7.811923 &    5.83 &  4.86 &       \\   &   27.234083 &    3.17 &  5.54 &       \\ \hline
084344 &    0.681833 &   19.44 &  5.40 &    f1 \\   &    0.103870 &   15.75 &  5.39 &       \\   &    0.343072 &   13.38 &  4.30 & 0.5f1 \\   &    0.025860 &   13.32 &  4.14 &       \\ \hline
085531 &    0.156886 &   30.31 &  8.04 &       \\ \hline
085719 &    0.060339 &   19.57 & 11.63 &       \\   &    0.026291 &   11.90 &  9.69 &       \\   &    0.086199 &    6.65 &  7.02 &       \\   &    0.113783 &    4.37 &  5.20 &       \\ \hline
086480 &    0.098267 &    3.22 &  5.22 &       \\   &    0.077148 &    2.65 &  4.77 &       \\ \hline
087084 &    1.076207 &   57.14 &  5.04 &    f1 \\   &    2.153707 &   46.23 &  4.85 &   2f1 \\   &    3.229483 &   38.13 &  4.92 &   3f1 \\   &    4.306552 &   35.92 &  6.69 &   4f1 \\   &    1.013281 &   27.29 &  4.42 &       \\   &    0.534440 &   23.17 &  3.27 & 0.5f1 \\   &    5.383621 &   24.95 &  5.88 &   5f1 \\   &    6.458535 &   18.87 &  6.12 &   6f1 \\ \hline
087501 &    5.168169 &   70.45 & 20.64 &    f1 \\   &   10.335475 &    8.13 &  6.08 &   2f1 \\   &    5.012145 &    5.68 &  4.97 &       \\ \hline
087548 &    5.057288 &   55.97 & 26.35 &    f1 \\   &   10.114145 &    6.42 & 14.67 &   2f1 \\   &    7.585932 &    3.24 &  9.01 &       \\   &   15.171002 &    3.03 &  8.37 &   3f1 \\   &   13.037149 &    1.71 &  5.40 &       \\ \hline
088142 &    6.826951 &  147.51 & 25.30 &    f1 \\   &   13.654333 &   36.48 & 19.73 &   2f1 \\   &    0.064649 &   29.18 &  4.53 &       \\   &    0.960256 &   29.40 &  5.02 &       \\   &   20.481714 &   13.42 & 10.30 &   3f1 \\   &    0.207308 &   22.77 &  4.40 & \\   &    3.413044 &   14.68 &  5.05 & 0.5f1 \\   &    1.018871 &   20.22 &  4.24 & \\   &    0.440045 &   18.50 &  3.90 & \\   &   10.240857 &    6.84 &  5.27 & 1.5f1 \\   &   12.034657 &    5.75 &  4.83 & \\ \hline
089391 &    7.305550 &    9.25 & 11.43 &    f1 \\   &    0.032756 &    4.93 &  4.89 &       \\   &    0.170247 &    3.33 &  3.44 &       \\   &   10.026622 &    2.91 &  4.19 &       \\   &   14.614248 &    2.75 &  3.64 &   2f1 \\ \hline
093873 &    0.029308 &   27.23 &  8.64 &       \\   &    0.061202 &   23.25 &  9.51 &       \\   &    0.018102 &   18.44 &  9.59 &       \\   &    0.052151 &   14.87 &  9.76 &       \\   &    0.076717 &    5.74 &  6.10 &       \\   &    0.092664 &    5.47 &  5.74 & \\   &    0.121110 &    2.51 &  4.90 & \\ \hline
094793 &    3.246138 &   10.17 &  4.60 &       \\ \hline
096404 &    1.718997 &  281.91 & 11.73 &    f1 \\   &    3.437994 &  152.42 & 14.47 &   2f1 \\   &    5.157423 &   92.84 & 13.79 &   3f1 \\   &    6.875988 &   63.18 & 12.98 &   4f1 \\   &    8.595848 &   37.14 & 10.38 &   5f1 \\   &   10.314413 &   24.19 &  7.21 &   6f1 \\   &   12.033411 &   17.12 &  6.25 &   7f1 \\ \hline
097790 &    1.917082 &    4.44 & 13.08 &       \\   &    1.874413 &    2.80 &  7.20 &       \\   &    1.858897 &    1.36 &  4.96 &       \\   &    2.017935 &    1.02 &  3.85 &       \\ \hline
098092 &    0.011206 &   58.09 & 16.75 &       \\   &    0.026291 &   34.54 & 13.86 &       \\   &    0.018533 &   15.87 &  9.68 &       \\ \hline
098719 &    4.802876 &  108.42 & 20.78 &    f1 \\   &    9.605320 &   22.83 & 18.49 &   2f1 \\   &   14.407334 &    7.77 & 12.48 &   3f1 \\ \hline
099529 &    0.206448 &    3.35 &  6.18 &       \\   &    0.043531 &    2.63 &  4.62 &       \\   &    0.117231 &    2.06 &  4.97 &       \\ \hline
106019 &    0.031036 &   83.13 &  9.22 &       \\   &    0.016811 &   29.97 &  7.61 &       \\ \hline
106769 &    0.040946 &   48.72 & 11.55 &       \\ \hline
107478 &    0.911567 &   42.45 &  7.09 &    f1 \\   &    1.823996 &   33.86 &  6.91 &   2f1 \\   &    2.735563 &   29.58 &  7.46 &   3f1 \\   &    3.647561 &   23.96 &  7.15 &   4f1 \\   &    0.453844 &   18.51 &  6.45 & 0.5f1 \\   &    1.368428 &   21.26 &  7.20 & 1.5f1 \\   &    2.280426 &   20.84 &  7.80 & 2.5f1 \\   &    3.192424 &   18.00 &  8.40 & 3.5f1 \\   &    4.559990 &   18.62 &  9.55 &   5f1 \\   &    4.103991 &   15.04 &  9.17 & 4.5f1 \\   &    5.471557 &   14.64 & 10.45 &   6f1 \\   &    5.015989 &   11.41 &  9.55 & 5.5f1 \\   &    6.383986 &    9.50 &  9.42 &   7f1 \\   &    0.462464 &   12.19 &  5.15 & \\   &    5.928418 &    8.44 & 10.24 & 6.5f1 \\   &    0.471946 &    7.60 &  5.10 & \\   &    0.444362 &    7.53 &  4.61 & \\   &    6.839554 &    5.84 &  9.18 & 7.5f1 \\   &    7.295984 &    4.87 &  9.14 &   8f1 \\ \hline
110801 &    0.028015 &   34.27 & 14.13 &       \\   &    0.011637 &   26.02 & 16.38 &       \\   &    0.039652 &    8.21 &  7.22 &       \\ \hline
111298 &    0.083182 &    6.18 &  7.77 &       \\   &    0.021981 &    5.45 &  7.12 &       \\   &    0.104301 &    5.37 &  7.22 &       \\ \hline
113453 &    0.028446 &   10.70 &  8.95 &       \\   &    0.074994 &    4.44 &  5.74 &       \\ \hline
114506 &    0.519781 &    8.08 &  6.39 &       \\   &    0.534435 &    4.15 &  4.47 &       \\ \hline
116410 &    0.271096 &    8.72 &  6.95 &       \\ \hline
116471 &    3.798428 &   27.83 &  5.84 &       \\ \hline
117654 &    0.046117 &   27.08 & 12.46 &       \\   &    0.058616 &    8.41 &  5.86 &       \\   &    0.029308 &    8.93 &  7.01 &       \\   &    0.011206 &    4.04 &  4.44 &       \\ \hline
118705 &    0.118095 &   13.83 &  8.44 &       \\ \hline
119488 &    0.012499 &   14.66 & 10.79 &       \\   &    0.023705 &   11.98 &  8.62 &       \\   &    0.031894 &   12.09 &  5.63 &       \\   &    0.054737 &    8.77 &  6.89 &       \\   &    0.062926 &    5.01 &  4.92 &       \\ \hline
120188 &    0.093527 &   10.44 &  9.82 &       \\   &    0.081459 &    5.43 &  7.74 &       \\   &    0.179726 &    3.37 &  4.91 &       \\   &    4.011735 &    2.30 &  7.53 &       \\ \hline
121369 &    0.047410 &   10.20 &  9.89 &       \\   &    0.012930 &    5.28 &  6.47 &       \\   &    0.065512 &    3.74 &  5.43 &       \\   &    0.080166 &    3.14 &  3.85 &       \\
121389 &    0.024577 &   37.09 &  8.41 &       \\   &    0.044842 &   30.04 &  4.98 &       \\   &    1.021869 &   17.90 &  4.20 &       \\ \hline
123934 &    8.195368 &   15.63 & 11.09 &    f1 \\   &    0.033619 &    7.85 &  4.96 &       \\   &    6.017436 &    5.62 &  4.43 &       \\   &    4.012774 &    4.97 &  3.97 &       \\   &   16.390305 &    4.40 &  4.00 &   2f1 \\ \hline
124666 &    2.183140 &  260.22 & 14.74 &    f1 \\   &    4.366280 &  106.12 & 15.01 &   2f1 \\   &    6.548989 &   58.61 & 14.30 &   3f1 \\   &    8.731267 &   28.31 & 10.96 &   4f1 \\   &   10.913976 &   17.01 &  9.17 &   5f1 \\   &    2.171933 &   16.61 &  4.51 &    f2 \\   &    2.191329 &   14.84 &  4.22 &    f3 \\   &    4.355936 &   13.15 &  5.33 &   2f2 \\   &    6.538213 &   11.71 &  6.46 &   3f2 \\   &   13.096254 &    9.98 &  7.99 &   6f1 \\   &    4.375762 &    8.86 &  4.12 &   2f3 \\   &   15.280687 &    7.11 &  7.00 &   7f1 \\   &    8.720106 &    6.64 &  4.94 &   4f2 \\   &   10.926521 &    6.57 &  5.09 &   5f3 \\   &   10.902021 &    5.66 &  4.91 &   5f2 \\   &   17.465235 &    5.05 &  5.22 &   8f1 \\   &   19.648375 &    4.90 &  5.38 &   9f1 \\   &    6.558585 &    5.41 &  3.80 &   3f3 \\   &   15.292439 &    4.73 &  4.94 &   7f3 \\ \hline
128178 &    0.063889 &   61.37 & 11.47 &       \\ \hline
133742 &    2.357630 &  201.87 & 12.96 &    f1 \\   &    4.714828 &   61.60 & 10.89 &   2f1 \\   &    7.072026 &   26.21 &  8.52 &   3f1 \\   &    1.178384 &   24.54 &  5.30 & 0.5f1 \\   &    3.536875 &   16.25 &  4.72 & 1.5f1 \\   &    9.430688 &    5.97 &  4.44 &   4f1 \\   &   13.029390 &    6.70 &  5.15 & 5.5f1
\enddata
\end{deluxetable}
\end{appendix}
\end{document}